\documentclass[12pt]{article}
\pdfoutput=1
\usepackage{amsmath,amssymb,amsfonts,epsfig,graphicx,euscript}%
\usepackage[pdftex,bookmarks,bookmarksnumbered,linktocpage,pdfstartview=FitH]{hyperref}
\hypersetup{colorlinks,%
citecolor=red,%
filecolor=blue,%
linkcolor=blue,%
urlcolor=blue,%
pdftex}
\usepackage{amsmath}
\usepackage{amsfonts}
\usepackage{graphicx}
\usepackage{caption}
\usepackage{subcaption}
\usepackage{float}
\usepackage[all]{hypcap}
\numberwithin{equation}{section}
\usepackage{cite}
\usepackage{calc}
\newlength{\spacer}
\newsavebox{\mybox}

\newcommand{\bse}{\begin{subequations}}
\newcommand{\ese}{\end{subequations}}
\newcommand{\be}{\begin{equation}}
\newcommand{\ee}{\end{equation}}
\newcommand{\bea}{\begin{eqnarray}}
\newcommand{\eea}{\end{eqnarray}}
\newcommand{\ba}{\begin{array}}
\newcommand{\ea}{\end{array}}

\begin{document}


\begin{center}
{ \large{\textbf{On the Contributions to the $\bf U_Y(1)$ Chern-Simons Term and the Evolution of Fermionic Asymmetries and Hypermagnetic Fields  }}} 
\vspace*{1.5cm}
\begin{center}
{\bf \small S. Rostam Zadeh\footnote{S$_{-}$Rostamzadeh@sbu.ac.ir}, S. S. Gousheh\footnote{ss${-}$gousheh@sbu.ac.ir}}\\%
\vspace*{0.4cm}
{\it {Department of Physics, Shahid Beheshti University, G.C., Evin, Tehran 19839, Iran}}  \\
\vspace*{0.3cm}
\begin{center}
\today
\end{center}
\end{center}
\end{center}

\bigskip
\begin{center}
\textbf{Abstract}
\end{center}
We study simultaneous evolution of electron, neutrino and quark asymmetries, and large scale hypermagnetic fields in the symmetric phase of the electroweak plasma in the temperature range $100$GeV$\leq T\leq 10$TeV, taking into account the chirality flip processes via inverse Higgs decays and fermion number violation due to Abelian anomalies.
We present a derivation of the coefficient of the Chern-Simons term for the hypercharge gauge field, showing that the left-handed and right-handed components of each fermion species contribute with opposite sign. This is in contrast to the results presented in some of the previous works. The $\textrm{U}_{\textrm{Y}}(1)$ Chern-Simons term affects the resulting anomalous magnetohydrodynamic (AMHD) equations. We solve the resulting coupled evolution equations for the lepton and baryon asymmetries, as well as the hypermagnetic field to obtain their time evolution along with their values at the electroweak phase transition ($T_{EW} \sim 100$GeV) for a variety of critical ranges for their initial values at $T=10$TeV. We first investigate the results of this sign change, by directly comparing our results with those obtained in one of the previous works and find that matter asymmetry generation increases considerably in the presence of a strong hypermagnetic field. Furthermore, we find that a strong hypermagnetic field can generate matter asymmetry starting from absolutely zero asymmetry, while matter asymmetry can generate a hypermagnetic field provided the initial value of the latter is nonzero.

\newpage

\tableofcontents

\section{Introduction}\label{Introduction}
The Universe in its hot early stages may have contained some magnetic fields. In fact, some large scale magnetic fields coherent over scales of the order of $30$ Kpc have been measured in our galaxy \cite{Kronberg}. The strength of these fields in the Milky Way and several spiral galaxies is of the order of the microgauss \cite{Kronberg}. It is believed that some primordial seed fields whose nature is largely unknown \cite{Kronberg, Kulsrud}, are needed for the generation of galactic magnetic fields \cite{Harrison}. Seed fields might be produced during the epoch of galaxy formation, or ejected by first supernova or active galactic nuclei \cite{Semikoz, Dvornikov}. Alternatively, they might arise from phase transitions in the very early universe \cite{Quashnock, Cheng, Kibble, Vachaspati, Enqvist, Baym} down to the cosmological inflation epoch \cite{Grasso}. New signatures for the presence of cosmological magnetic fields (CMF) in the intergalactic medium and their survival till today has been observed recently \cite{Vovk, Neronov}. 

At high temperatures, the non-Abelian interactions induce a \textit{magnetic} mass gap $\sim$ $g^2T$ for their corresponding gauge fields. As a consequence, non-Abelian magnetic fields (corresponding to the color SU(3) or weak SU(2) groups) cannot survive and long range fields of these types do not exist. Therefore, the only long range field surviving in the plasma is associated with the Abelian U(1) group \cite{Kajantie} since its vector particle remains massless. Moreover, Abelian and non-Abelian electric fields quickly decay because of the finite plasma conductivity. The large scale hypercharge magnetic field existing in the symmetric phase of the primordial plasma is converted to the ordinary Maxwellian CMF during the electroweak symmetry breaking \cite{Giovannini}.

Hypermagnetic fields interact with matter differently from what ordinary magnetic fields do. The hypercharge fields couple to the fermions chirally, while the coupling of the ordinary electromagnetic fields is vector-like. Consequently, the simultaneous presence of hyperelectric and hypermagnetic fields leads to the fermion number violation due to the Abelian anomaly, $\partial_\mu j^\mu \sim \frac{g'^2}{4\pi^2} \textbf{E}_\textbf{Y}.\textbf{B}_\textbf{Y}$. Here, $g'$ is the $\textrm{U}_{\textrm{Y}}(1)$ gauge coupling \cite{Giovannini}. The Abelian anomaly states that the hypercharge fields are coupled to the fermionic number density. This effect also appears as the Chern-Simons term for the corresponding gauge fields.

It was realized long ago that a Chern-Simons term appears to be induced in the effective Lagrangian density describing the dynamics of the SU(2) gauge fields at high temperatures and finite fermionic densities \cite{Redlich}. Since the Abelian hypercharge fields couple to the chiral fermions in the symmetric phase, a Chern-Simons term emerges for them as well \cite{Giovannini, Laine}. The hypermagnetic Chern-Simons term gives rise to an anomalous term in the magnetohydrodynamic (MHD) equations governing the evolution of the hypercharge magnetic fields in the high-temperature electroweak plasma. The resulting extended equations are called the Anomalous MHD (AMHD) equations \cite{Giovannini, Joyce}. To summarize, 
a hypermagnetic source term appears in the evolution equations of the anomalous charge densities (Abelian anomaly), and the anomalous term shows up as the Chern-Simons term in the Lagrangian for the hypermagnetic field and affects its evolution equation. These Anomalous couplings could generate instabilities, leading to the exchange of energy between fermionic sector and magnetic hypercharge fields \cite{Giovannini, Laine, Joyce, Rubakov}. This might have important effects in cosmology.

The electroweak plasma in complete thermal equilibrium can be described by $n_G$ chemical potentials ${\mu}_i$, $i=1,..., n_G$ related to $n_G$ strictly conserved global charges $N_i = B/n_G -L_i$. Here, $B$ is the baryon number, $L_i$ is the lepton number of the $i$-th generation and $n_G$ is the number of generations. Moreover, another chemical potential $\mu_Y$ corresponding to weak hypercharge is introduced which will be fixed as a consequence of the hypercharge neutrality of the plasma, $\langle Y \rangle = 0$ \cite{Gorbunov}.

It was shown years ago that perturbative reactions with right-handed electron chirality flip are out of thermal equilibrium at temperatures higher than $T_{RL} \sim 10$ TeV \cite{Campbell}. This is due to the fact that their Yukawa coupling with Higgs bosons $h_e = 2.94 \times {10}^{-6}$ is very small so the rates ${\Gamma}_{RL} \sim h_e^2T$ of the related processes\footnote{There are some gauge and fermion scattering processes (such as $e_R H\leftrightarrow L_e A$, where $A=Y$ or $W$, and $e_R L_f\leftrightarrow L_e f_R$) in addition to the direct and inverse Higgs decays, which participate in the chirality flip of electrons and have contributions to the total chirality flip rate  
(see the third paper of \cite{Campbell}).
} (direct and inverse Higgs decays in reactions $e_L \bar{e}_R\leftrightarrow\phi^{(0)}$ and $\nu_e^L \bar{e}_R\leftrightarrow\phi^{(+)}$ and their conjugate reactions) are much lower than the Hubble expansion rate $H \sim T^2$ in this range of temperatures. As a consequence, the number of right-handed electrons is perturbatively conserved for temperatures $T>T_{RL}$ unless Abelian anomaly ($\partial_\mu j_{e_R}^\mu = \frac{g'^2}{4\pi^2} \textbf{E}_\textbf{Y}.\textbf{B}_\textbf{Y}$) is taken into account. 
Thus, the electroweak theory acquires an extra partially conserved charge and its corresponding chemical potential is added to the mentioned $n_G + 1 = 4$ (for 3 generations) chemical potentials. There also exist some relations between these $n_G + 2 = 5$ chemical potentials and their asymmetries (see Eq.\ (2.16) of \cite{Giovannini}). 

In a scenario suggested by the authors of \cite{Campbell}, a primordial baryon asymmetry is preserved by an asymmetry in the number of right-handed electrons 
which are protected from washing out by weak sphalerons down to $T_{RL}$\footnote{The value of $T_{RL}$ computed in the first paper of Ref.\ \cite{Campbell} was $T_{RL}\simeq 1$TeV.}.  
Since, the weak sphalerons start to fall out of thermal equilibrium at roughly the same temperature\footnote{In recent years, the temperature at which the weak sphalerons start to fall out of thermal equilibrium has been computed more accurately \cite{Burnier}.},
it is possible that the transformation of right-handed electrons into left-handed leptons is not very significant during the overlap time of these two processes. Thus, the weak sphalerons may not be able to turn them into antiquarks and thereby wash out the remaining baryon and lepton asymmetries \cite{Campbell}.

Afterwards in a related work, the authors of \cite{Giovannini} accounted for the Abelian anomaly for right-handed electrons and also assumed that a large scale hypermagnetic field is present in the plasma. 
They also supposed that left-handed leptons have no asymmetry. Therefore, weak sphaleron processes were not considered. 
In addition, they assumed that no Higgs asymmetry and hence no net contribution to chirality flip processes from direct Higgs decays is present \cite{Dvornikov}. However, chirality flip processes due to inverse Higgs decay were taken into account. They solved the evolution equations for the right-handed electron asymmetry and the hypermagnetic field in the adiabatic approximation analytically. In continuation, the authors of \cite{Dvornikov2011} analyzed the same physical problem beyond the adiabatic regime and solved the dynamical equations numerically. Then, they broadened the scenario by assuming that lepton asymmetries for left-handed electrons and electron neutrinos ($\mu_{e_L} = \mu_{\nu_e^L}$) are also present but the weak sphaleron processes and Higgs asymmetry are still absent. In a subsequent work, these authors took into account the presence of Higgs asymmetry and weak sphalerons as well \cite{Dvornikov}. In recent studies, they investigated the evolutions by assuming a continuous helicity spectrum for the hypermagnetic field \cite{Smirnov,Sokoloff}.


The reverse situation has been studied by considering the existence of an asymmetry for right-handed electrons and the Abelian anomaly which leads to the appearance of a hypermagnetic Chern-Simons term. When the full range of frequency spectrum of the hypermagnetic field is taken into account, this term gives rise to instabilities and generation of the long range hypermagnetic fields \cite{Joyce} which could affect the electroweak phase transition \cite{Elmfors}, the sphaleron energy \cite{Comelli} and the electroweak baryogenesis \cite{Piccinelli}.

The main purpose of this paper is to present one correction to the form of the $\textrm{U}_{\textrm{Y}}(1)$ Chern-Simons term which is sometimes used in the literature, and investigate its important effects on the evolution of fermionic asymmetries and hypermagnetic field. We will show that the sign of chemical potential for right-handed particles appearing in this Chern-Simons term is opposite to that of the left-handed ones, contrary to what has been used in some of the previous works \cite{Dvornikov2011,Dvornikov,Smirnov,Sokoloff}. 
We investigate and show the effect of this correction on the evolutions by comparing our results with those of Ref.\ \cite{Dvornikov2011} in particular. For this purpose, our basic assumptions have to be the same as theirs. 
In particular, we have neglected processes such as direct Higgs decays, and weak sphaleron processes and all of their consequences in our model. The scope of our work is described below. 

In this work we study both cases, that is, the conversion of initial hypermagnetic field into fermion asymmetries, and vice versa, in the same temperature range as considered in \cite{Dvornikov2011,Dvornikov,Smirnov,Sokoloff}, i.e. $T_{EW}\sim100$GeV$\leq T \leq10$TeV\footnote{It has  been shown recently that the onset of electroweak phase transition is at $T\simeq160$GeV \cite{Burnier}. However in this study we adhere to the usual practice of using $T_{EW}\sim 100$GeV.}. As a matter of fact, our investigation is one step more thorough in the sense that we solve the set of coupled differential equations for a variety of ranges of initial conditions for both the fermion asymmetries and the hypermagnetic field. 
We consider fermion number violation due to the Abelian anomalies for electrons, neutrinos and quarks in the presence of hypermagnetic fields. In addition, the electron chirality flip reactions through inverse Higgs decays are accounted for\footnote{For the reason just mentioned, we have neglected extra chirality flip processes mentioned in the footnote on page 4. Nevertheless, we did re-run our programs after including the extra chirality flip processes, to study their effects. We observed that the two chiral components followed each other even more closely especially in Figures \ref{one-}, \ref{three-timeplot} and \ref{three-}. However, the general features of the graphs were not altered and the new graphs are not displayed in this paper.
}, since they violate chiral lepton numbers and enter into thermal equilibrium below $T_{RL}$.
That is, these terms tend to reduce any initial fermionic chiral asymmetry present at $T_{RL}$, as $T$ decreases. The value of electron chiral asymmetry $\Delta\mu = \mu_{e_R}-\mu_{e_L}$ before EWPT is important because it has been shown that the evolution of Maxwellian magnetic fields in the broken phase at temperatures $10$ MeV $<T<T_{EW}$ is highly affected by this parameter since their evolutions are strongly coupled \cite{Boyarsky}.



The outline of the paper is the following. In Section \ref{Static Chern-Simons Terms}, we use the effective Euclidean Lagrangian of the gauge fields at finite fermionic density to present a derivation of the Chern-Simons coefficients in terms of the chemical potentials of right-handed and left-handed particle species. In Section \ref{Anomalous MHD Equations}, we use the flat space effective Lagrangian for the hypercharge field $Y_\mu$ containing the hypermagnetic Chern-Simons term of Section \ref{Static Chern-Simons Terms}, to derive the dynamical equations for hypercharge fields. Then, we combine these equations to get the evolution equation of long range hypermagnetic field. In Section \ref{Kinetics Of Leptons In Hypermagnetic Fields}, we derive the dynamical equations for the asymmetries of right-handed and left-handed electrons (positrons) accounting for the Abelian anomalies and the inverse Higgs decays. In Section \ref{Baryon Asymmetry Generation In Hypermagnetic Fields}, we derive the analogous evolution equation of the baryon asymmetries. In Section \ref{Discussion}, we solve numerically the set of coupled differential equations for the hypermagnetic field and the asymmetries of baryons and first generation leptons in the symmetric phase, starting at $10$TeV and ending at the onset of the electroweak phase transition (EWPT) $T_{EW}\sim100$GeV, and present the results. We use the important conservation law $B/3-L_e=\mbox{constant}$ merely as a consistency check on our results. We also use the conventions discussed in Appendix A and the anomaly equations summarized in Appendix B of Ref. \cite{Long}. In Section \ref{Summary and discussion}, we summarize our main results and present our concluding remarks.
 
\section{Static Chern-Simons Terms}\label{Static Chern-Simons Terms}
As mentioned in Section \ref{Introduction}, the Chern-Simons terms are induced in the effective action of the $\textrm{SU}(2)_{\textrm{L}}$ and $\textrm{U}_{\textrm{Y}}(1)$ gauge fields in the presence of fermionic chemical potentials. The effective action for the soft gauge fields in the static limit can be derived by using the method of Dimensional Reduction \cite{Ginsparg}. In the standard electroweak theory at temperatures $T$ above a few hundred GeV, the gauge field part of the dimensionally reduced Euclidean Lagrangian \cite{Kajantie1995} takes the form \cite{Laine}
\be\begin{split}\label{Lagrangian}
{\cal{L}}_E = f_E + \frac{1}{4} G_{\mu\nu}^a G_{\mu\nu}^a + \frac{1}{4} Y_{\mu\nu} Y_{\mu\nu} +  \frac{1}{2} m_E^2 A_0^a A_0^a + \frac{1}{2} m_E^{'2} Y_0 Y_0 + \cr
+ i j'_E Y_0 + c_E n_{CS} + c'_E n'_{CS} + ...\ \ \ \ \ \ \ \ \ \ \ \ \ \ \ \ \ \ \ \ \ \ \ \ \ \ \ \ \ \ \ \                         
\end{split}\ee 
where the Chern-Simons densities are \cite{Laine}
\be
n_{CS} = \frac{g^2}{32\pi^2} {\epsilon}_{ijk} (A_i^a G_{jk}^a - \frac{g}{3} f^{abc} A_i^a A_j^b A_k^c)\ ,
\ee
\be
n'_{CS} = \frac{g'^2}{32\pi^2} {\epsilon}_{ijk} Y_i Y_{jk}.\ \ \ \ \ \ \ \ \ \ \ \ \ \ \ \ \ \ \ \ \ \ \ \ \ \ \ 
\ee
In the above equations $A_{\mu}^a$ and $Y_{\mu}$ are the $\textrm{SU}(2)_{\textrm{L}}$ and $\textrm{U}_{\textrm{Y}}(1)$ gauge fields, and $G_{\mu\nu}^a$, $Y_{\mu\nu}$, $g$ and $g'$ are the corresponding field strength tensors and gauge couplings, respectively.

The Chern-Simons coefficients $c_E$ and $c'_E$ are given by Eqs.\ (2.8) and (2.9) of Ref.\ \cite{Laine}, assuming that all quarks have the same chemical potential. Below, we present the general form of these coefficients without the above assumption. Since the non-Abelian gauge interactions are in thermal equilibrium at all temperatures of interest, they enforce equality of the asymmetries carried by different components of a given multiplet \cite{Long}. Thus, we can let $\mu_{Q_i}$ denote the common chemical potential of up and down left-handed quarks with different colors, $\mu_{{u_R}_i}$($\mu_{{d_R}_i}$) the common chemical potential of right-handed up (down) quarks with different colors and $\mu_{L_i}$ ($\mu_{R_i}$) the common chemical potential of left-handed (right-handed) leptons, where \lq{\textit{i}}\rq\ is the generation index. The coefficient $c_E$ can now be written as \cite{Redlich}, 
\be\label{c_E}
c_E = {\sum}_{i=1}^{n_G}(3\mu_{Q_i}+{\mu}_{L_i}).
\ee

The coefficient $c'_E$ can be derived using the prescription given in the footnote on page 5 of Ref.\ \cite{Laine}, which states that a single chiral fermion with the chemical potential $\mu$ and the hypercharge $Y$ contributes to the parameter $c'_E$ via the equation $c'_E = -\mu Y^2 H/2$. Here, $H$ is $+1$ ($-$1) for right-handed (left-handed) fermions. Thus, $c'_E$  for all fermion species can be written as  
\be\begin{split}\label{c'_E1}
c'_E = {\sum}_{i=1}^{n_G} [-\mu_{R_i} Y_{R}^2 (\frac{1}{2}) - \mu_{L_i} Y_{L}^2 (\frac{-1}{2}) N_w -\mu_{{d_R}_i} Y_{d_R}^2 (\frac{1}{2})N_c +\cr
-\mu_{{u_R}_i} Y_{u_R}^2 (\frac{1}{2})N_c - \mu_{Q_i} Y_{Q}^2 (\frac{-1}{2}) N_w N_c ],\ \ \ \ \ \ \ \ \ \ \ \ \ \  
\end{split}\ee
where $N_c = 3$ and $N_w = 2$ are the ranks of the non-Abelian gauge groups, and the hypercharges are \cite{Long}
\be 
Y_Q = \frac{1}{3},\ \ \ Y_{u_R} = \frac{4}{3},\ \ \ Y_{d_R} = -\frac{2}{3},\ \ \ Y_L = -1,\ \ \ Y_R = -2.
\ee
Substituting the above constants in Eq.\ (\ref{c'_E1}) we get
\be\begin{split}\label{c'_E2}
c'_E = {\sum}_{i=1}^{n_G} \left[-2\mu_{R_i} + \mu_{L_i} - \frac{2}{3}\mu_{{d_R}_i} - \frac{8}{3}\mu_{{u_R}_i} + \frac{1}{3}\mu_{Q_i} \right],\ \ \ \ \ \ \ \ \ \  
\end{split}\ee
which is the coefficient of the static hypermagnetic Chern-Simons term taking into account the chemical potentials of all leptons and quarks\footnote{Sometimes in literature the phrase `back reaction' is used for the contributions of various chemical potentials to $c'_E$, perhaps implying a second order effect. However, as we shall show, in our model the mutual effects of the chemical potentials and the hypermagnetic field on each other are of the same order.}.

In some previous works, various models for the evolution of matter asymmetries and hypermagnetic field have been considered in which the only contributions taken into account for $c'_E$ as given by Eq.\ (\ref{c'_E2}), have been due to chemical potentials of leptonic asymmetries of the first generation \cite{Dvornikov2011,Dvornikov,Smirnov,Sokoloff}. In that case the expression for $c'_E$ reduces to  $-2\mu_{e_R} + \mu_{e_L}$, in contrast to \cite{Dvornikov2011,Dvornikov,Smirnov,Sokoloff} where it is taken as $-2\mu_{e_R} - \mu_{e_L}$. It is precisely the consequences of this relative sign difference that we want to explore in this paper. In order to accomplish this task, we choose one of these models, i.e. the simple model of Ref.\ \cite{Dvornikov2011}, and use the same main assumptions, aside from the aforementioned sign difference.

Although the weak sphalerons are not taken into account in the simple model that we want to study and have just described, it is worthwhile to briefly describe some of their consequences. Let us first concentrate on the well studied cases where the hypermagnetic field is absent. The rate of weak sphaleron processes is much higher than the Hubble expansion rate in the whole symmetric phase \cite{Burnier}. Therefore, as the weak sphalerons approach chemical equilibrium very rapidly, they force $c_E$ to vanish (see Table 1 of Ref.\ \cite{Long}). Let us now include the hypermagnetic field and first assume for simplicity that the effect of weak sphalerons on $c_E$ is unaffected. Then considering the usual simplifying assumptions for the chemical potentials ($\mu_{{d_R}_i}=\mu_{{u_R}_i}=\mu_{Q_i}=\mu_{Q};\ i=1,2,3$, and $\mu_{R_i}=\mu_{L_i};\ i=2,3$), the expression for $c'_E$ reduces to $c'_E \simeq -2\mu_{e_R} + 2\mu_{e_L}-c_E$. One can then neglect $c_E$ as compared to $-2\mu_{e_R} + 2\mu_{e_L}$ when the weak sphalerons are in equilibrium,  and the expression for $c'_E$ simplifies to $c'_E \simeq -2\mu_{e_R} + 2\mu_{e_L}$. However, note that the relative sign difference still remains\footnote{However, it should be noted that merely setting $c_E=0$ as described above is not sufficient to take the weak sphalerons into account. To have a consistent set of evolution equations one has to include the corresponding term for the weak sphalerons in the evolution equations of left-handed fermionic asymmetries and let $c_E$ evolve freely in accordance to the evolutions of its constituents as given by Eq.\ (\ref{c_E}).}. When we include properly both the effects of the weak sphaleron processes and the hypermagnetic field, we find that the two terms are in competition. Although the sphaleron processes are usually the dominant effect and $c_E$ stays very close to zero, we find that as we get close to $T_{EW}$, the effect of the hypermagnetic field becomes strong enough to force the system out of equilibrium. We also find other interesting effects that deserve further investigation and we plan to report on them elsewhere. Now let us return to our model.

The conversion of Eq.\ (\ref{Lagrangian}) to Minkowski spacetime is accomplished by analytic continuation \cite{Laine}
\be
\partial_0^E = -i\partial_0^M,\ \ \ A_0^{aE} = -i A_0^{aM},\ \ \ B_0^E = -iB_0^M ,\ \ \ {\cal{L}}_E = -{\cal{L}}_M\ .   
\ee 
Thus, the hypermagnetic Chern-Simons term in the effective Minkowskian Lagrangian is given by
\be
\Delta{\cal{L}}_M = -\Delta{\cal{L}}_E = -c'_E n'_{CS} = -c'_E\frac{g'^2}{32\pi^2}(2\textbf{Y}.\textbf{B}_\textbf{Y})\ .
\ee
In the above equation, $\textbf{B}_\textbf{Y}$ = $\nabla\times\textbf{Y}$ is the hypermagnetic field.

\section{Anomalous MHD Equations}\label{Anomalous MHD Equations}

Let us assume that the electroweak plasma is slightly out of thermal equilibrium due to the presence of the large scale hypermagnetic fields or some unbalanced chemical potentials. As mentioned in section \ref{Introduction}, the hypercharge fields are coupled to the fermionic number densities because of the anomaly. In this paper, we set up  
and solve the coupled system of Boltzmann-type equations for these chemical potentials and the hypermagnetic fields. In this section, we derive the dynamical equations of the hypercharge fields. 

The dynamical equations for hypermagnetic and hyperelectric fields, taking into account the anomaly (AMHD equations), can be derived using the flat space effective Lagrangian for the hypercharge field $Y_\mu$ at finite fermion density obtained in Section \ref{Static Chern-Simons Terms},
\be
{\cal{L}} = -\frac{1}{4} Y_{\mu\nu} Y^{\mu\nu} - J_{Y}^\mu Y_\mu - c'_E\frac{g'^2}{32\pi^2}(2\textbf{Y}.\textbf{B}_\textbf{Y}),
\ee
where $J_\mu = (J_{Y}^0,\textbf{J}_\textbf{Y})$ is the vector Ohmic current with zero time component due to the hypercharge-neutrality of the plasma $J_{Y}^0 = \langle Y \rangle = 0$. The evolution equations are :
\be\begin{split}\label{AMHD} 
\frac{\partial \textbf{B}_\textbf{Y}} {\partial t} = - \nabla\times\textbf{E}_\textbf{Y},\ \ \ \ \ \frac{\partial \textbf{E}_\textbf{Y}} {\partial t} + \textbf{J}_\textbf{Y} - c'_E\frac{g'^2}{8\pi^2}\textbf{B}_\textbf{Y} = \nabla\times\textbf{B}_\textbf{Y},\cr
\nabla.\ \textbf{B}_\textbf{Y} = 0,\ \ \ \ \ \ \ \ \ \ \ \ \ \ \ \  \ \ \ \ \nabla.\ \textbf{E}_\textbf{Y} = 0,\cr
\nabla.\ \textbf{J}_\textbf{Y} = 0,\ \ \ \ \ \ \ \ \ \ \ \ \ \  \textbf{J}_\textbf{Y} = \sigma(\textbf{E}_\textbf{Y} + \textbf{V}\times\textbf{B}_\textbf{Y}).
\end{split}\ee
For the hot electroweak plasma, the hyperconductivity is $\sigma \sim 100T$ \cite{Arnold}.

As can be seen in Eq.\ (\ref{AMHD}), a new \lq{fermionic}\rq\ current ($-c'_E\frac{g'^2}{8\pi^2}\textbf{B}_\textbf{Y}$) is induced due to the presence of finite chemical potentials. This is the hypercharge current which is a kind of Ohmic current flowing parallel to the hypermagnetic field. It is important to note that both the current and the hypermagnetic field are vector fields and the chemical potentials of the anomalous charges appear in the proportionality factor \cite{Giovannini2013}. 

It is worth mentioning that the anomaly term appearing in the above equations for the hypercharge field was first introduced in the context of chiral magnetic effect in Ref.\ \cite{Vilenkin}. 
It was argued that if the Hamiltonian of a system of charged fermions does not conserve parity, then an equilibrium electric current can develop in such a system parallel to an external magnetic field $\textbf{B}$ \cite{Vilenkin}. To understand this effect, let us investigate one particular effect of an external Maxwellian magnetic field on right-handed and left-handed charged fermions and antifermions. The magnetic field couples to the magnetic moments of the particles and tends to align them. Therefore, the spins of particles which have positive (negative) electric charge are aligned (anti-aligned) with the magnetic field. Then, the helicity eigenvalues of the particles specify the direction of their momentum, from which the direction of the electric current corresponding to each particle is determined. Assuming that the four helicity states of these particles are massless, the net electric current for electrons will be $\textbf{J}\propto[n(e^-_R)-n(e^+_L)]-[n(e^-_L)-n(e^+_R)]$, 
where $n(e^-_R)$ is the number density of right-handed electrons and similarly for the other particle species. The terms in the brackets are given by the chemical potentials of right-handed and left-handed electrons respectively, namely $\textbf{J}\propto\mu_5=\mu_{e^-_R}-\mu_{e^-_L}$. The current is in the direction of the magnetic field $\textbf{B}$, and based on the calculations of Ref.\ \cite{Vilenkin} is in the form $\textbf{J}=(e^2/2\pi^2)\mu_5 \textbf{B}$. This is the current emerging when chiral charge and magnetic field are present in the plasma. In the symmetric phase, a similar expression for the hypercharge current can be obtained in which, the magnetic field is replaced by the hypermagnetic field and $\mu_5$ by its analogue in the symmetric phase (see Eq.\ (4.24) of Ref.\ \cite{Long2016}).

The generalized Ohm law at finite hyperconductivity can be written as 
\be 
\textbf{E}_\textbf{Y} = \frac{\textbf{J}_\textbf{Y}}{\sigma} - \textbf{V}\times\textbf{B}_\textbf{Y} = \frac{1}{\sigma}(\nabla\times\textbf{B}_\textbf{Y} + c'_E\frac{g'^2}{8\pi^2}\textbf{B}_\textbf{Y} - \frac{\partial \textbf{E}_\textbf{Y}} {\partial t} )  - \textbf{V}\times\textbf{B}_\textbf{Y}.
\ee
To be consistent with the standard MHD approach, the displacement current ${\partial \textbf{E}_\textbf{Y}}/{\partial t}$ which is sub-leading, provided the conductivity is finite, can be neglected. As a consequence, the hyperelectric field can be written as
\be \label{E_Y}
\textbf{E}_\textbf{Y} = - \textbf{V}\times\textbf{B}_\textbf{Y} + \frac{\nabla\times\textbf{B}_\textbf{Y}}{\sigma} - \alpha_Y\textbf{B}_\textbf{Y},\ \ \ \mbox{where}\ \ \ 
\alpha_Y(T) = - c'_E\frac{g'^2}{8\pi^2\sigma}. \ \ \ \ \ \  
\ee
The hypermagnetic helicity coefficient $\alpha_Y$ originates from the $\textrm{U}_{\textrm{Y}}(1)$ Chern-Simons term and is a scalar.

Now, by inserting $\textbf{E}_\textbf{Y}$ from Eq.\ (\ref{E_Y}) into the evolution equation (\ref{AMHD}) of the hypermagnetic field, the generalized magnetic diffusivity equation can be obtained:
\be
\frac{\partial \textbf{B}_\textbf{Y}} {\partial t} = \nabla\times(\textbf{V}\times\textbf{B}_\textbf{Y}) + \frac{1}{\sigma}\nabla^2\textbf{B}_\textbf{Y} + \alpha_Y\nabla\times\textbf{B}_\textbf{Y}.
\ee

We make a further simplification and assume that the electroweak plasma is globally parity-invariant, thus no global vorticity is present. Now, since the correlation distance of the hypermagnetic field is much larger than the length scale of the bulk velocity field variation, the hypercharge infrared modes   are not practically affected by the plasma velocity. Therefore, concerning the large-scale part of the hypercharge, the velocity field will be neglected \cite{Giovannini} and the evolution equation for $B_Y$ reduces to
\be \label{hypermagnetic}
\frac{\partial \textbf{B}_\textbf{Y}} {\partial t} = \frac{1}{\sigma}\nabla^2\textbf{B}_\textbf{Y} + \alpha_Y\nabla\times\textbf{B}_\textbf{Y},
\ee 
where $\alpha_Y$ contains time varying chemical potentials and is obtained by Eqs.\ (\ref{c'_E2},\ref{E_Y}).

\section{The Evolution Equations for the Lepton Asymmetries and the Hypermagnetic Field }\label{Kinetics Of Leptons In Hypermagnetic Fields}

To the equation of motion for the hypermagnetic field (\ref{hypermagnetic}), we should add the evolution equations of the leptonic and baryonic chemical potentials. In this section, we obtain the equations for leptonic asymmetries by assuming that there is no Higgs boson asymmetry while taking into account the Abelian anomalous contributions and inverse Higgs decay processes.

The lepton number violation due to the Abelian anomaly is given by \cite{Giovannini, Dvornikov2011},
\be\label{e_R} 
\partial_\mu j_{e_R}^\mu = -\frac{1}{4}(Y_R^2) \frac{g'^2}{16\pi^2}Y_{\mu\nu} {\tilde{Y}}^{\mu\nu} = \frac{g'^2}{4\pi^2}(\textbf{E}_\textbf{Y}.\textbf{B}_\textbf{Y}),
\ee
\be \label{e_L}
\partial_\mu j_{\nu_e^L}^\mu = \partial_\mu j_{e_L}^\mu = +\frac{1}{4}(Y_L^2) \frac{g'^2}{16\pi^2}Y_{\mu\nu} {\tilde{Y}}^{\mu\nu} = - \frac{g'^2}{16\pi^2}(\textbf{E}_\textbf{Y}.\textbf{B}_\textbf{Y}).
\ee
As a consequence, the system of dynamical equations for leptons accounting for the Abelian anomalies (\ref{e_R}),(\ref{e_L}) and perturbative chirality flip reactions takes the form \cite{Dvornikov2011,Dvornikov}, 
\be\begin{split}\label{lepton equations} 
\frac{d\eta_{e_R}}{dt} = \frac{g'^2}{4\pi^2s}(\textbf{E}_\textbf{Y}.\textbf{B}_\textbf{Y}) + 2\Gamma_{RL}(\eta_{e_L}-\eta_{e_R}),\cr
\frac{d\eta_{e_L}}{dt} = -\frac{g'^2}{16\pi^2s}(\textbf{E}_\textbf{Y}.\textbf{B}_\textbf{Y}) + \Gamma_{RL}(\eta_{e_R}-\eta_{e_L}),\cr
\frac{d\eta_{\nu_e^L}}{dt} = -\frac{g'^2}{16\pi^2s}(\textbf{E}_\textbf{Y}.\textbf{B}_\textbf{Y}) + \Gamma_{RL}(\eta_{e_R}-\eta_{\nu_e^L}).
\end{split}\ee
In the above equations, $\eta_b = (n_b -n_{\bar{b}})/s$ with $b=\{e_R,e_L,\nu_e^L\}$ is the lepton asymmetry, $s=2\pi^2g^*T^3/45$ is the entropy density and $g^*=106.75$ is the number of relativistic degrees of freedom. Assuming that the SU(2) gauge interactions are rapid, we can use the approximation $\eta_{e_L}\approx \eta_{\nu_e^L}$. The factor 2 multiplying the rate $\Gamma_{RL}$ in the first line is due to the equivalent rates of inverse Higgs decay processes.
 The rate of these processes $\Gamma_{RL}$ is \cite{Campbell,Dvornikov2011,Dvornikov}
\be 
\Gamma_{RL} = 5.3\times 10^{-3}h_e^2(\frac{m_0}{T})^2T = (\frac{\Gamma_0}{2t_{EW}})(\frac{1-x}{\sqrt{x}}),
\ee
where the variable $x = t/t_{EW} = (T_{EW}/T )^2$ is given by the Friedmann law, $t_{EW} = M_0/2T_{EW}^2$ and $M_0 = M_{Pl}/1.66\sqrt{g^*}$. Moreover, $h_e = 2.94 \times {10}^{-6}$ is the Yukawa coupling for electrons, $\Gamma_0 = 121$ and $m_0^2(T) = 2DT^2(1 -T_{EW}^2/T^2)$ is the temperature dependent effective Higgs mass at zero momentum and zero Higgs vacuum expectation value. The coefficient $2D \sim 0.377$ in the expression for $m_0^2(T)$ has contributions coming from the known masses of gauge bosons $m_Z$ and $m_W$, the top quark mass $m_t$, and the zero-temperature Higgs mass. (see Ref. \cite{Aad}). The rate $\Gamma_{RL}$ vanishes at the EWPT time $x = 1$.

Let us recall the equation $n_b -n_{\bar{b}} = \mu_b T^2/6$ and define $\xi_b = \mu_b/T$. Then, the lepton asymmetry will be $\eta_b = \xi_b T^3/6s$. Now, Eqs.\ (\ref{lepton equations}) can be written in terms of the asymmetries $\xi_b$ as
\be\begin{split}\label{asymmetry equations} 
\frac{d\xi_{e_R}}{dt} = \frac{3g'^2}{2\pi^2T^3}(\textbf{E}_\textbf{Y}.\textbf{B}_\textbf{Y}) + 2\Gamma_{RL}(\xi_{e_L}-\xi_{e_R}),\cr
\frac{d\xi_{e_L}}{dt} = -\frac{3g'^2}{8\pi^2T^3}(\textbf{E}_\textbf{Y}.\textbf{B}_\textbf{Y}) + \Gamma_{RL}(\xi_{e_R}-\xi_{e_L}),\cr
\frac{d\xi_{\nu_e^L}}{dt} = -\frac{3g'^2}{8\pi^2T^3}(\textbf{E}_\textbf{Y}.\textbf{B}_\textbf{Y}) + \Gamma_{RL}(\xi_{e_R}-\xi_{e_L}).
\end{split}\ee
Since $\xi_{\nu_e^L}=\xi_{e_L}$, the third equation for neutrinos is superfluous and there are only two independent equations for two chemical potentials.

We can simplify the form of the Abelian anomaly contribution in the equations ($\sim \textbf{E}_\textbf{Y}.\textbf{B}_\textbf{Y}$) by choosing the simplest nontrivial configuration of the hypermagnetic field, which is
\be\label{wave}
Y_x=Y(t)\sin k_0z,\ \ \ \ \ Y_y=Y(t)\cos k_0z,\ \ \ \ \ Y_z =Y_0 =0.
\ee
Substituting $\textbf{E}_\textbf{Y}$ from Eq.\ (\ref{E_Y}) into $\textbf{E}_\textbf{Y}.\textbf{B}_\textbf{Y}$ leads to
\be\label{E_Y.B_Y1} 
\textbf{E}_\textbf{Y}.\textbf{B}_\textbf{Y} = \frac{1}{\sigma}(\nabla\times\textbf{B}_\textbf{Y}).\textbf{B}_\textbf{Y} - \alpha_Y\textbf{B}_\textbf{Y}^2.
\ee
For the simple choice of wave configuration we obtain, $(\nabla\times\textbf{B}_\textbf{Y}).\textbf{B}_\textbf{Y}=k_0B_Y^2(t)$ and $\textbf{B}_\textbf{Y}^2=B_Y^2(t)$, where $B_Y(t)=k_0Y(t)$ is the hypermagnetic field amplitude. Thus, (\ref{E_Y.B_Y1}) becomes
\be\label{E_Y.B_Y2} 
\textbf{E}_\textbf{Y}.\textbf{B}_\textbf{Y} = \frac{1}{\sigma}(k_0B_Y^2) - \alpha_Y B_Y^2.
\ee 
Now, we substitute $c'_E$ from Eq.\ (\ref{c'_E2}) into the expression for $\alpha_Y$ as given by Eq.\ (\ref{E_Y}) and 
neglect the contribution of 
all chemical potentials except $\mu_{e_R},\mu_{e_L}$ and $\mu_{\nu_e^L}$, in order to confine the calculations to the model under study and be able to compare our results with those of Ref.\ \cite{Dvornikov2011}.
Then, we obtain 
\be\begin{split}\label{alpha_Y} 
\alpha_Y(T) = - c'_E\frac{g'^2}{8\pi^2\sigma}\simeq (2\mu_{e_R} - \mu_{e_L})\frac{g'^2}{8\pi^2\sigma}.
\end{split}\ee
Putting the above $\alpha_Y$ into Eq.\ (\ref{E_Y.B_Y2}) and using $\sigma = 100T$ as assumed in \cite{Dvornikov2011,Dvornikov}, we obtain
\be\label{E_Y.B_Y3}
\textbf{E}_\textbf{Y}.\textbf{B}_\textbf{Y} = \frac{B_Y^2}{100} \left[\frac{k_0}{T}-\frac{g'^2}{4\pi^2}(\xi_{e_R}-\frac{\xi_{e_L}}{2})\right].
\ee
Using (\ref{E_Y.B_Y3}) and defining $y_R(x) = 10^4\xi_{e_R}(x)$ and $y_L(x) = 10^4\xi_{e_L}(x)$, Eqs.\ (\ref{asymmetry equations}) can be rewritten in the form 

\be\begin{split}\label{y_equations}
\frac{dy_R}{dx} = \left[B_0 x^{1/2}-A_0(y_R-\frac{y_L}{2})\right](\frac{B_Y(x)}{10^{20}\textrm{G}})^2x^{3/2} -\Gamma_0\frac{1-x}{\sqrt{x}}(y_R-y_L),\cr
\frac{dy_L}{dx} = -\frac{1}{4}\left[B_0 x^{1/2}-A_0(y_R-\frac{y_L}{2})\right](\frac{B_Y(x)}{10^{20}\textrm{G}})^2x^{3/2} -\Gamma_0\frac{1-x}{2\sqrt{x}}(y_L-y_R),
\end{split}\ee
where,
\be\label{A0B0}
B_0 = 25.6 (\frac{k_0}{10^{-7}T_{EW}}),\ \ \ \ \ A_0 = 77.6.
\ee
The overall scale of $B_0$ and $A_0$ is chosen so that the hypermagnetic field is normalized at $10^{20}\textrm{G}$. It should be noted that the constants $A_0$ and $B_0$ introduced in Ref.\ \cite{Dvornikov2011} are one fourth of the above ones. However, in Ref.\ \cite{Dvornikov} they are corrected to the values presented in Eq.\ (\ref{A0B0}). As a matter of fact the values that we obtain are $B_0=25.97$ and $A_0=82.68$, which differ slightly from those of Ref.\ \cite{Dvornikov}. However, we continue using the values as given in Eq.\ (\ref{A0B0}).

Substituting the simple wave configuration given in Eq.\ (\ref{wave}) into the equation of motion of the hypermagnetic field (\ref{hypermagnetic}), we obtain the evolution equations for $B_Y(t)$ in the form
\be
\frac{dB_Y}{dt} = B_Y(t)\left[-\frac{k_0^2}{\sigma(t)}+k_0\alpha_Y(t)\right]. 
\ee
Substituting $\alpha_Y(t)$ from (\ref{alpha_Y}) into the above equation results in,
\be \label{B_Y(t)}
\frac{dB_Y}{dt} = B_Y(t)\left[-\frac{k_0^2}{\sigma(t)}+\frac{k_0 g'^2}{4\pi^2\sigma(t)}(\mu_{e_R} - \frac{\mu_{e_L}}{2})\right].
\ee
Rewriting Eq.\ (\ref{B_Y(t)}) in terms of the previously defined parameters $x,y_R$ and $y_L$ leads to
\be \label{B_Y(x)}
\frac{dB_Y}{dx} = 3.5(\frac{k_0}{10^{-7}T_{EW}})\left[\frac{y_R-y_L/2}{\pi}-0.1(\frac{k_0}{10^{-7}T_{EW}})\sqrt{x})\right]B_Y(x).
\ee

The initial conditions for our first investigation chosen at $x_0 = 10^{-4}$ or at $T_0 = 10$TeV, 
are
\be
 y_R^{(0)}=y_R(x_0) = 10^{-6},\ \ \ \ \ y_L^{(0)}=y_L(x_0) = 0.
\ee
Such conditions correspond to the right-handed electron asymmetry $\xi_{e_R}(x_0) = 10^{-10}$  \cite{Dvornikov2011,Dvornikov}.

\section{The Evolution Equations for the Baryon Asymmetries}\label{Baryon Asymmetry Generation In Hypermagnetic Fields}
Baryogenesis and leptogenesis can arise in the presence of the hypermagnetic field fulfilling the condition $\eta_B/3 - \eta_{L_e} =\mbox{constant}$, where $\eta_B = (n_B-n_{\bar{B}})/s$ and $\eta_{L_e} = (n_{L_e}-n_{\bar{L}_e})/s$ are the baryon and the first generation lepton asymmetries respectively. Such baryogenesis can occur, since the hypermagnetic fields affect the baryon asymmetry similar to the lepton asymmetry due to the Abelian anomalies \cite{Long}:

\begin{eqnarray}\label{Q_equations}
\partial_\mu j_{Q_i}^\mu &= \frac{1}{4}(N_c N_w Y_Q^2) \frac{g'^2}{16\pi^2} Y_{\mu\nu} {\tilde{Y}}^{\mu\nu} &= -\frac{g'^2}{24\pi^2}(\textbf{E}_\textbf{Y}.\textbf{B}_\textbf{Y}),\\*\nonumber
\partial_\mu j_{u_{R_i}}^\mu &= -\frac{1}{4}(N_c Y_{u_R}^2) \frac{g'^2}{16\pi^2} Y_{\mu\nu} {\tilde{Y}}^{\mu\nu} &= \frac{g'^2}{3\pi^2}(\textbf{E}_\textbf{Y}.\textbf{B}_\textbf{Y}),\\*\nonumber
\partial_\mu j_{d_{R_i}}^\mu &= -\frac{1}{4}(N_c Y_{d_R}^2) \frac{g'^2}{16\pi^2} Y_{\mu\nu} {\tilde{Y}}^{\mu\nu} &= \frac{g'^2}{12\pi^2}(\textbf{E}_\textbf{Y}.\textbf{B}_\textbf{Y}).
\end{eqnarray}

Here suppressed gauge group indices are summed, but the generation index \lq{\textit{i}}\rq\ is not summed. Then, we get 
\be  
\partial_\mu j_B^\mu = \frac{1}{N_c}\sum_i(\partial_\mu j_{Q_i}^\mu + \partial_\mu j_{u_{R_i}}^\mu + \partial_\mu j_{d_{R_i}}^\mu) = \frac{3g'^2}{8\pi^2}(\textbf{E}_\textbf{Y}.\textbf{B}_\textbf{Y}),
\ee
and thus,
\be \label{baryon equation}
\frac{d\eta_B}{dt} = \frac{3g'^2}{8\pi^2s}(\textbf{E}_\textbf{Y}.\textbf{B}_\textbf{Y}).
\ee
Combining the three parts of Eq.\ (\ref{lepton equations}) and comparing the result with Eq.\ (\ref{baryon equation}), we obtain
\be
\frac{1}{3}\frac{d\eta_B}{dt} = \frac{d\eta_{e_R}}{dt} + \frac{d\eta_{e_L}}{dt} + \frac{d\eta_{\nu_e^L}}{dt} = \frac{d\eta_{L_e}}{dt},
\ee
which is precisely the expected conservation law $\eta_B/3 - \eta_{L_e} = \mbox{constant}$.

Defining $y_B(x) = 10^4\xi_{B}(x)$, where $\xi_{B} = \mu_B/T$, the above equation can be rewritten in the form 
\be 
\frac{1}{3}\frac{dy_B}{dx} = \frac{dy_R}{dx}+ 2\frac{dy_L}{dx}, 
\ee
and finally using Eq.\ (\ref{y_equations}), we obtain
\be\label{yB_equation}
\frac{dy_B}{dx} = \frac{3}{2}\left[B_0 x^{1/2}-A_0(y_R-\frac{y_L}{2})\right](\frac{B_Y(x)}{10^{20}G})^2x^{3/2}.
\ee
It should be noted that $\eta_B=(n_B-n_{\bar{B}})/s=\mu_BT^2/6s=\xi_BT^3/6s=10^{-4} y_B T^3/6s=45 \times 10^{-4} y_B/12\pi^2 g^*$.

\section{Results}\label{Discussion}
\begin{figure} 
  \includegraphics[width=65mm]{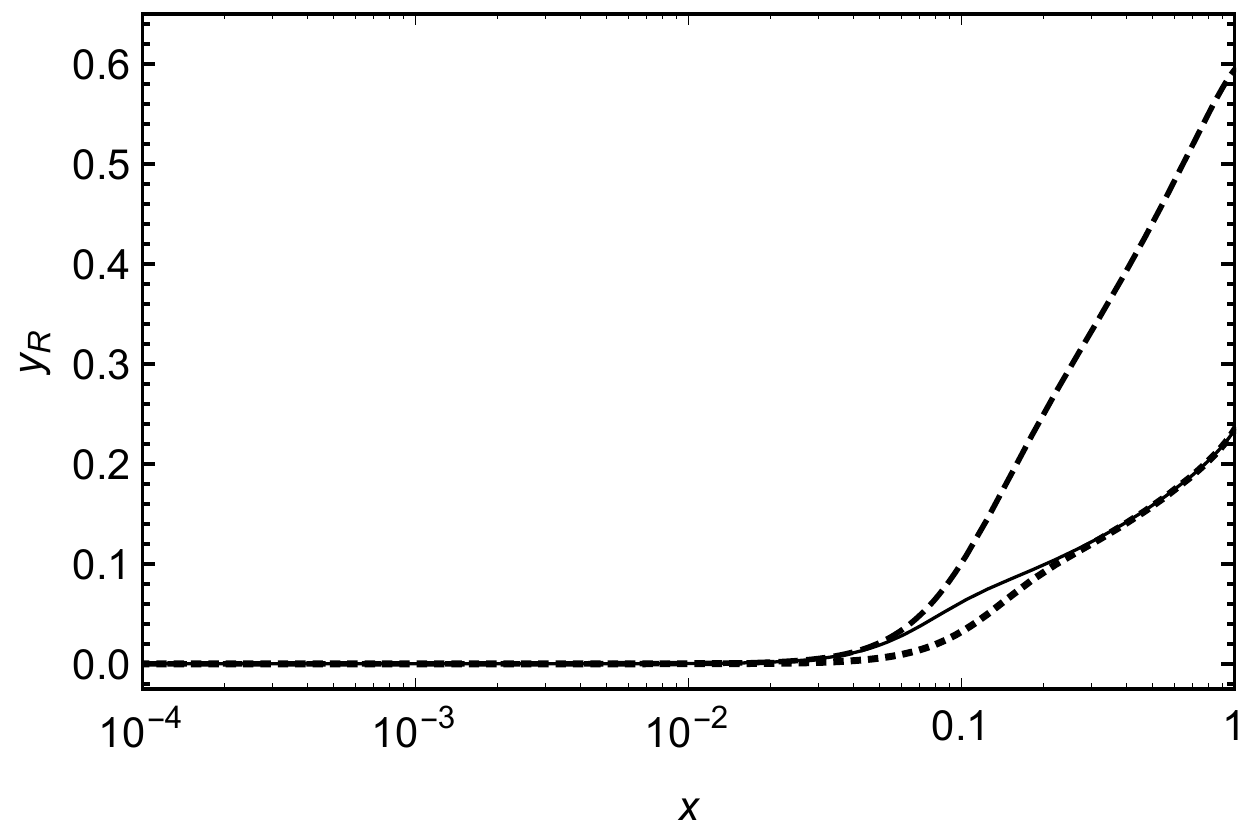}
  \hspace{2mm}
  \includegraphics[width=65mm]{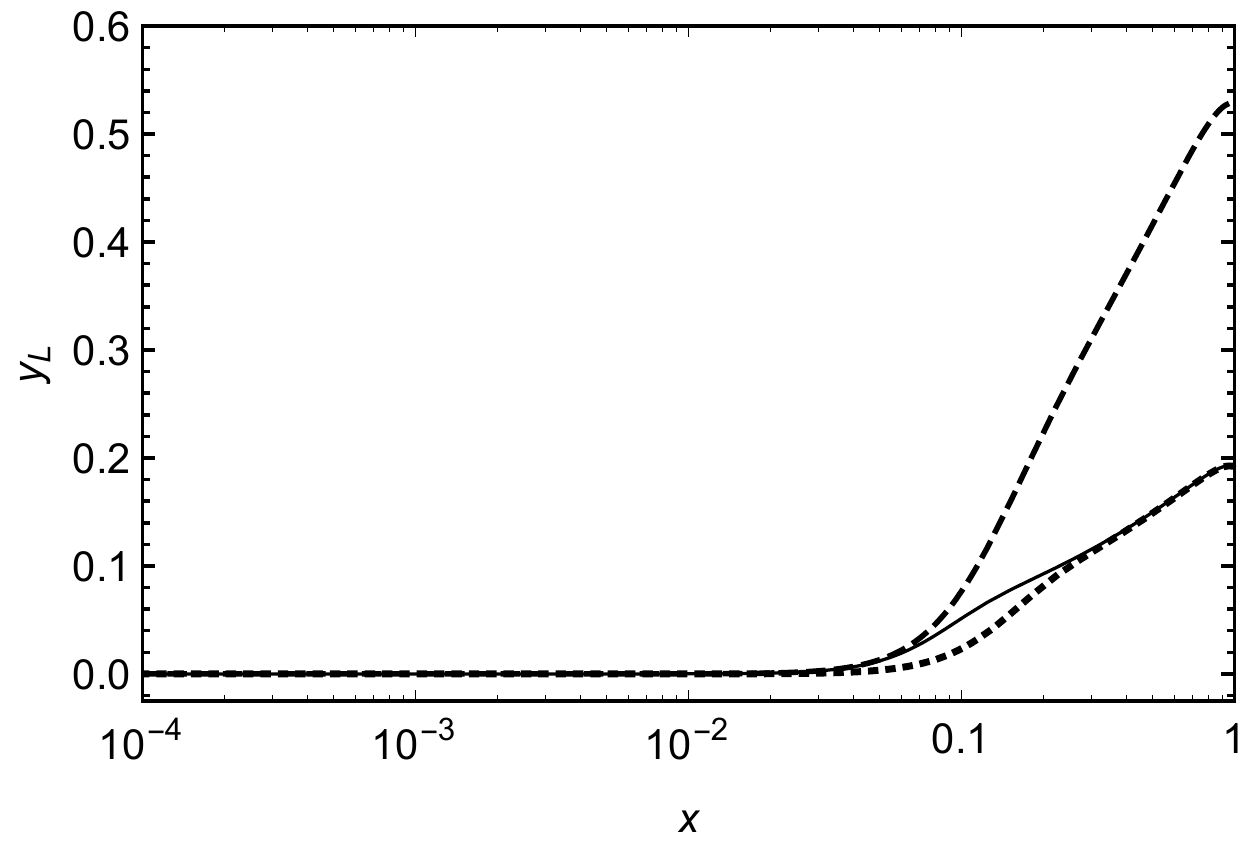}
  \includegraphics[width=65mm]{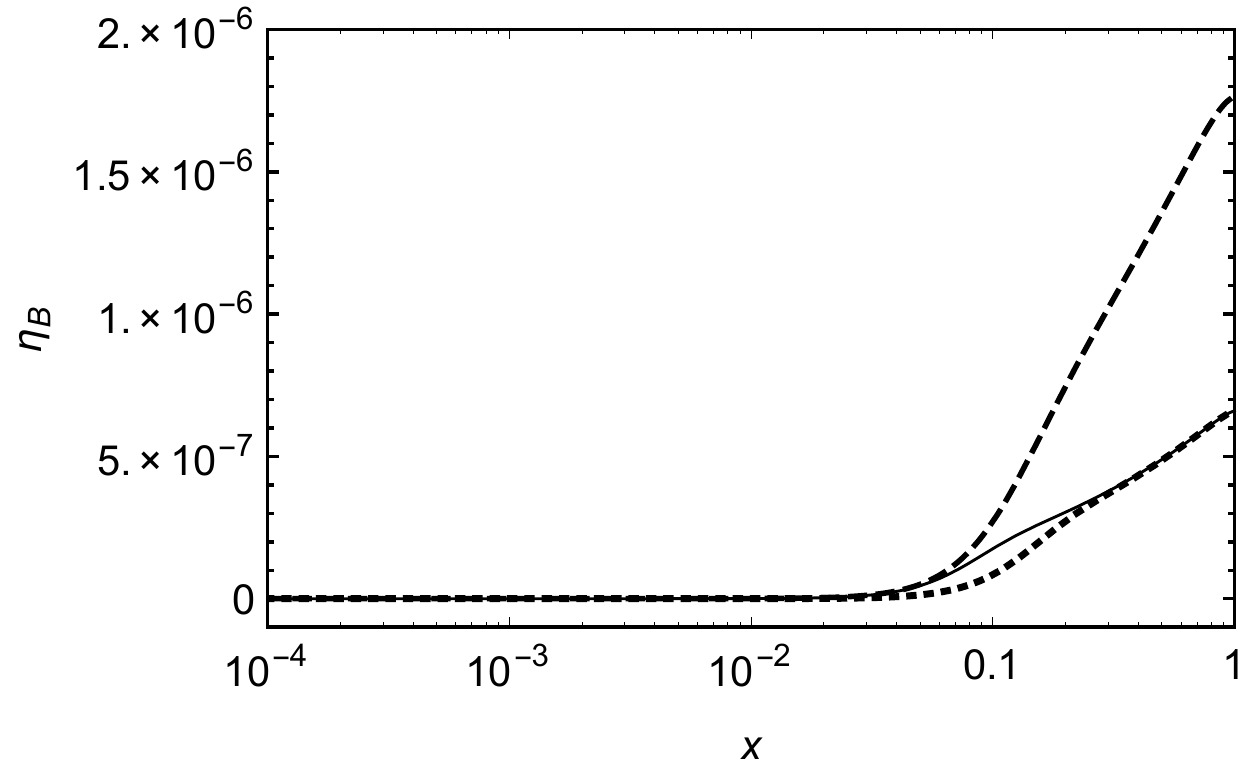}
  \hspace{2mm}
  \includegraphics[width=65mm]{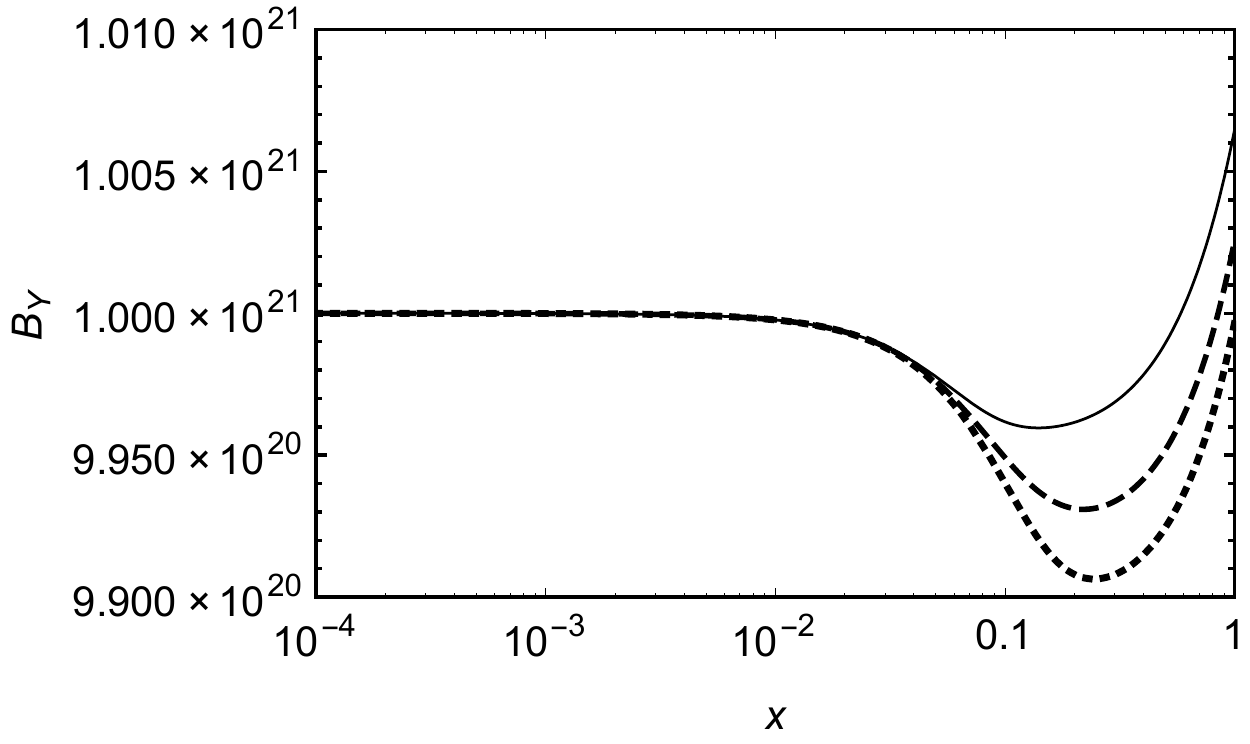}
\caption{The time plots of the normalized leptonic asymmetries $y_R$ and $y_L$, baryonic asymmetry $\eta_B$ and the hypermagnetic field amplitude $B_Y$ for $k_0=10^{-7} T_{EW}$ and initial conditions $y_R^{(0)} = 10^{-6}$ and $B_Y^{(0)}=10^{21}G$ for three different cases.\\
 Case 1 (dotted lines): \ \ \  $A_0 = 19.4$, $B_0 = 6.4$ and $\alpha_Y=\frac{g'^2}{8\pi^2\sigma}(2\mu_{e_R} + \mu_{e_L})$.\\ Case 2 (solid lines): \ \ \ \ \  $A_0 = 77.6$, $B_0 = 25.6$ and $\alpha_Y=\frac{g'^2}{8\pi^2\sigma}(2\mu_{e_R} + \mu_{e_L})$.\\ Case 3 (dashed lines):\ \ \ \ $A_0 = 77.6$, $B_0 = 25.6$ and $\alpha_Y = \frac{g'^2}{8\pi^2\sigma}(2\mu_{e_R} - \mu_{e_L})$.\\
For these time plots, the starting point is at $x_0=\frac{t_0}{t_{EW}}=(\frac{T_{EW}}{T_0})^2=10^{-4}$ and the final point is at the onset of the electroweak phase transition $x_f=\frac{t_f}{t_{EW}}=(\frac{T_{EW}}{T_f})^2=1$. Cases 1 and 2 are obtained from the assumptions of Refs.\ \cite{Dvornikov2011,Dvornikov} and are reproduced here for comparison purposes. Case 3 is the result of this paper.
The maximum relative error for these plots is of the order of $10^{-21}$.
}\label{RGB}

\end{figure}




To study the evolution of the fermionic asymmetries and the hypermagnetic field, we solve the set of four coupled differential equations simultaneously, as an initial value problem starting at $T=10$TeV ($x_0=10^{-4}$) and ending at $T=100$GeV$\sim T_{EW}$ ($x_f=1$). We assume a simple wave configuration for the hypermagnetic field as given in Eq.\ (\ref{wave}). Since the evolution equations for the asymmetries of left-handed electrons and neutrinos are the same, this set includes two equations for the right-handed and left-handed electrons (\ref{y_equations}), one equation for the baryons (\ref{yB_equation}) and one for the hypermagnetic field (\ref{B_Y(x)}). Because of the conservation of $\eta_B/3-\eta_{L_e}$ during the evolution, its value has been used as a check for the accuracy of the solutions. This is in contrast to the usual practice where this constraint is used to eliminate the evolution equation of baryons \cite{Dvornikov, Giovannini, Dvornikov2011}. 

As stated earlier, our main assertion is that the left-handed and right-handed fermion contributions to the Chern-Simons term for the hypercharge fields have opposite signs. This difference shows up in Eq.\ (\ref{c'_E1}) and affects the AMHD equations through the hypermagnetic helicity coefficient $\alpha_Y$ as given by Eqs.\ (\ref{E_Y},\ref{alpha_Y}). That is, we believe that the correct form of $\alpha_Y$ for the simple model that we study in this paper is $\alpha_Y= (2\mu_{e_R} - \mu_{e_L})\frac{g'^2}{8\pi^2\sigma}$. In this section we investigate the results of our evolution equations for a variety of initial conditions.

Our first investigation is focused on exploring the effect of our proposed form for $\alpha_Y$ as compared to the one used in other works such as Ref.\  \cite{Dvornikov2011}, i.e. $\alpha_Y= (2\mu_{e_R} - \mu_{e_L})\frac{g'^2}{8\pi^2\sigma}$ versus $\alpha_Y= (2\mu_{e_R} + \mu_{e_L})\frac{g'^2}{8\pi^2\sigma}$, respectively.
We use the same initial values $y_R^{(0)} = y_R(x_0) = 10^{-6}$, $B_Y^{(0)}=B_Y(x_0)=10^{21}G$ and the maximum wave number surviving Ohmic dissipation $k_0=10^{-7} T_{EW}$, and solve the equations numerically and present the results for three different cases in Figure \ref{RGB}. In each part of Figure \ref{RGB} we compare the effects of changing the values of $A_0$ and $B_0$, as well as the form of 
the hypermagnetic helicity coefficient $\alpha_Y$. 
It can be seen that putting the correct values of the  parameters $A_0$ and $B_0$ has little effect on the fermionic asymmetries, however the amplitude of hypermagnetic field increases slightly. More importantly, correcting the parameter $\alpha_Y$ as described above leads to an increase in the fermionic asymmetries, but smaller increase in the hypermagnetic field amplitude as compared to the second case.
For the rest of our analysis in this section, we use the corrected values of $A_0,B_0$ and corrected form of $\alpha_Y$ as described above, while $k_0$ is set to $10^{-7} T_{EW}$.

\begin{figure} 
  \includegraphics[width=65mm]{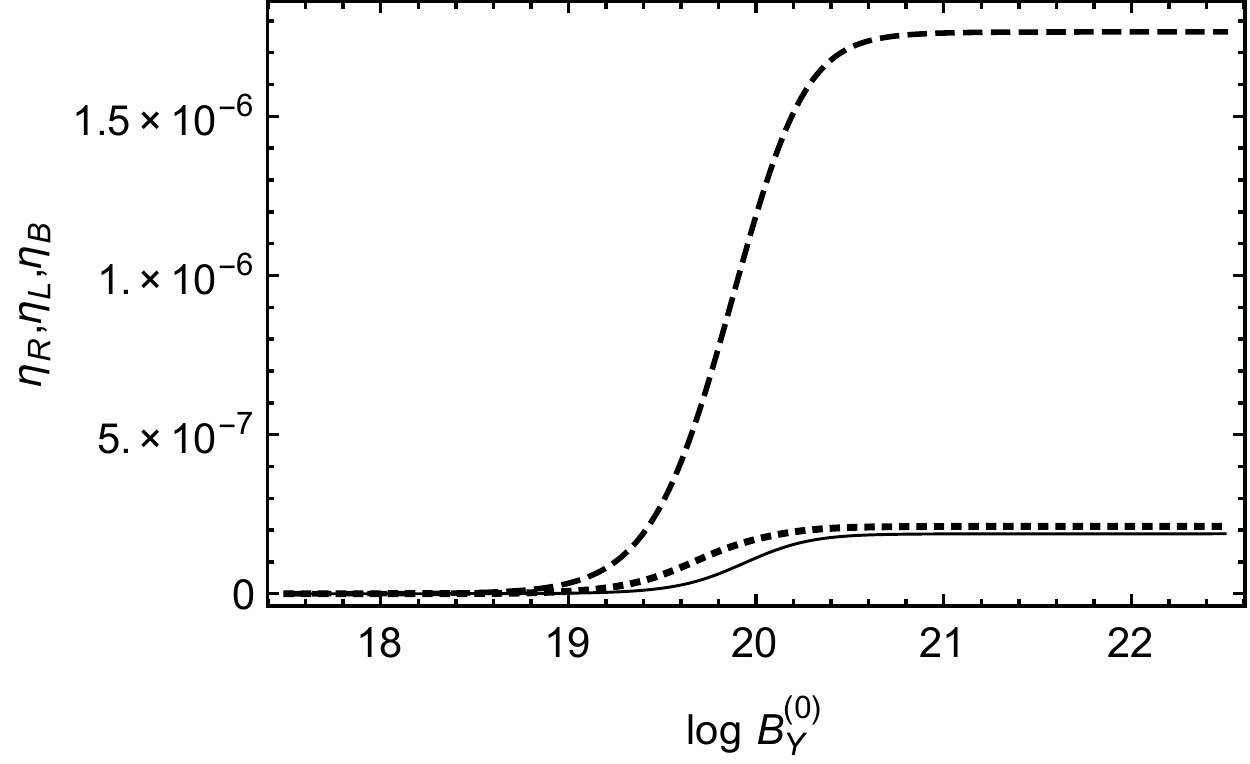}
  \hspace{2mm}
  \includegraphics[width=65mm]{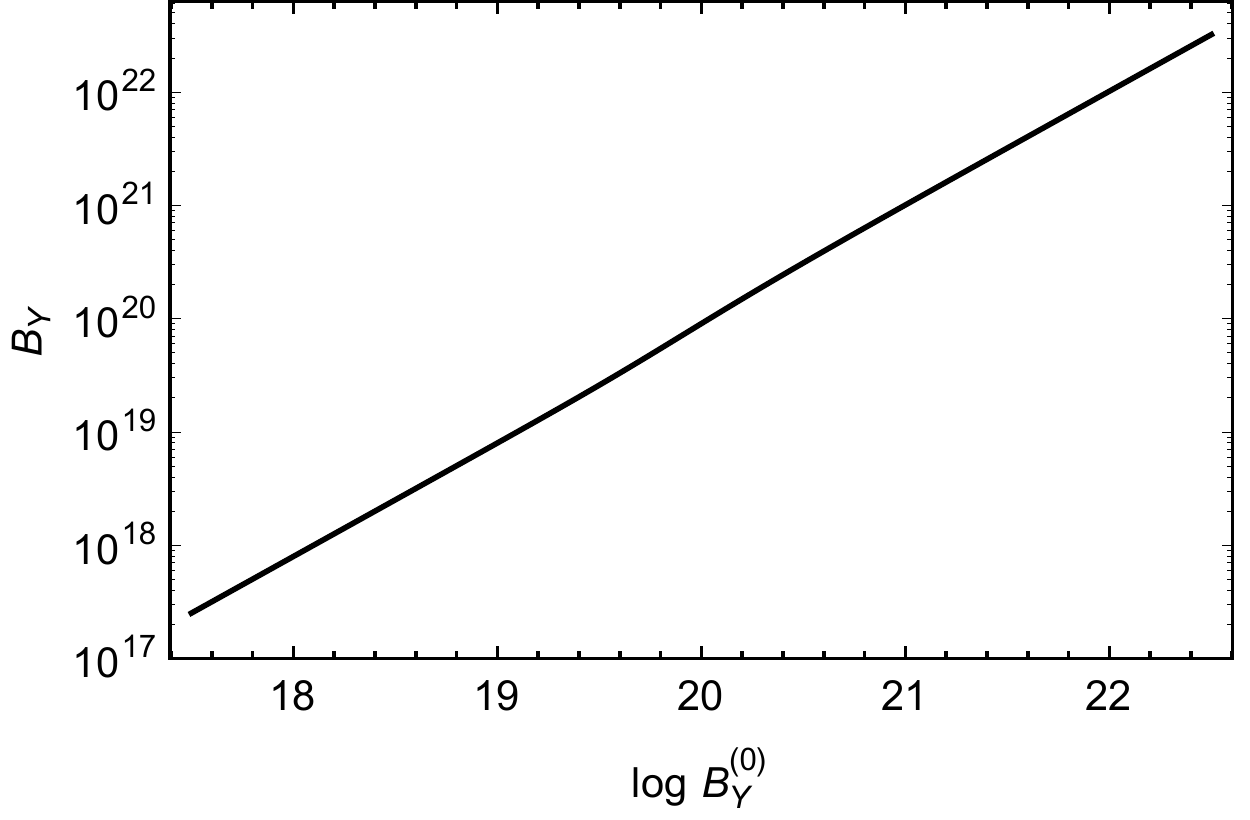}
\caption{Left: The asymmetries of right-handed electrons $\eta_R=\eta_{e_R}$ (dotted line), left-handed leptons $\eta_L=\eta_{e_L}=\eta_{\nu_e^L}$ (solid line) and baryons (dashed line) at the EWPT time $t_{EW}$ where x=1. Right: The amplitude of hypermagnetic field at the EWPT time $t_{EW}$. It is assumed that all of the initial fermionic asymmetries are zero, $k_0=10^{-7}T_{EW}$ and $B_Y^{(0)}$ changes between $10^{17.5}$G and $10^{22.5}$G. The maximum relative error for these graphs is of the order of $10^{-17}$.}\label{one-}

\end{figure}

In the second part of our investigation, we solve the evolution equations setting the initial values of all fermionic asymmetries, including that of right-handed electrons, to zero. However, we assume that the initial hypermagnetic field is nonzero. We solve the evolution equations for $10^{17.5}$G$\ <B_Y^{(0)}<10^{22.5}$G, and display the results at $T=T_{EW}$ in Figure \ref{one-}.
The interesting point is that, the asymmetries grow due to the presence of hypermagnetic fields even in the absence of any initial fermion asymmetry. Moreover, for $B_Y^{(0)}\geq 10^{21}$G, the hypermagnetic field amplitude grows very slightly (above its initial value) as well, which is the sign for a kind of weak resonance effect. The time plots are similar to those of case three in Figure \ref{RGB}, so we do not show them here. The left plot shows that the final matter asymmetries grow by increasing $B_Y^{(0)}$ especially between $10^{19}$G and $10^{20.5}$G, but remain nearly constant when $B_Y^{(0)}$ becomes larger than $10^{21}$G. The inflection point occurs near $B_Y^{(0)}=10^{20}$G. However, in all cases the final value of $B_Y$ is nearly the same as its initial value. It can be seen from the right plot that $B_Y{(t_{EW})}$ also has an inflection point, though very very mild, near the critical point $B_Y^{(0)}=10^{20}$G. Another interesting conclusion from these graphs is that $B_Y^{(0)}=10^{20.5}$G is the most effective value for generating matter asymmetries.  
 

In the third part of our analysis, we investigate the reverse situation. Namely, we assume that the initial value of the hypermagnetic field amplitude is zero but there exists some primordial right-handed electron asymmetry. We solve the equations for $y_R^{(0)} = 10^3$ and $y_L^{(0)} = 0$. We do not display the graphs, as they can be easily described. We observe that as time elapses the right-handed electron asymmetry is transformed into the left-handed lepton asymmetry due to the chirality flip processes until they become equal. Moreover, the baryonic asymmetry remains zero and net leptonic asymmetry does not change because the hypermagnetic field amplitude stands at zero. This is due to the fact that, for the simple wave configuration of the hypermagnetic field, the dynamical equation for the field amplitude is (\ref{B_Y(x)}) which can be integrated to yield
\be
B_Y(x)=B_Y^{(0)}\exp\left[\frac{3.5k_0}{10^{-7}T_{EW}} \int_{x_0}^x (\frac{y_R(x')-y_L(x')/2}{\pi}-\frac{0.1k_0}{10^{-7}T_{EW}}\sqrt{x'})dx'\right]. 
\ee
It is clear that the hypermagnetic field freezes at zero if its initial value is exactly zero.

\begin{figure} 
  \includegraphics[width=65mm]{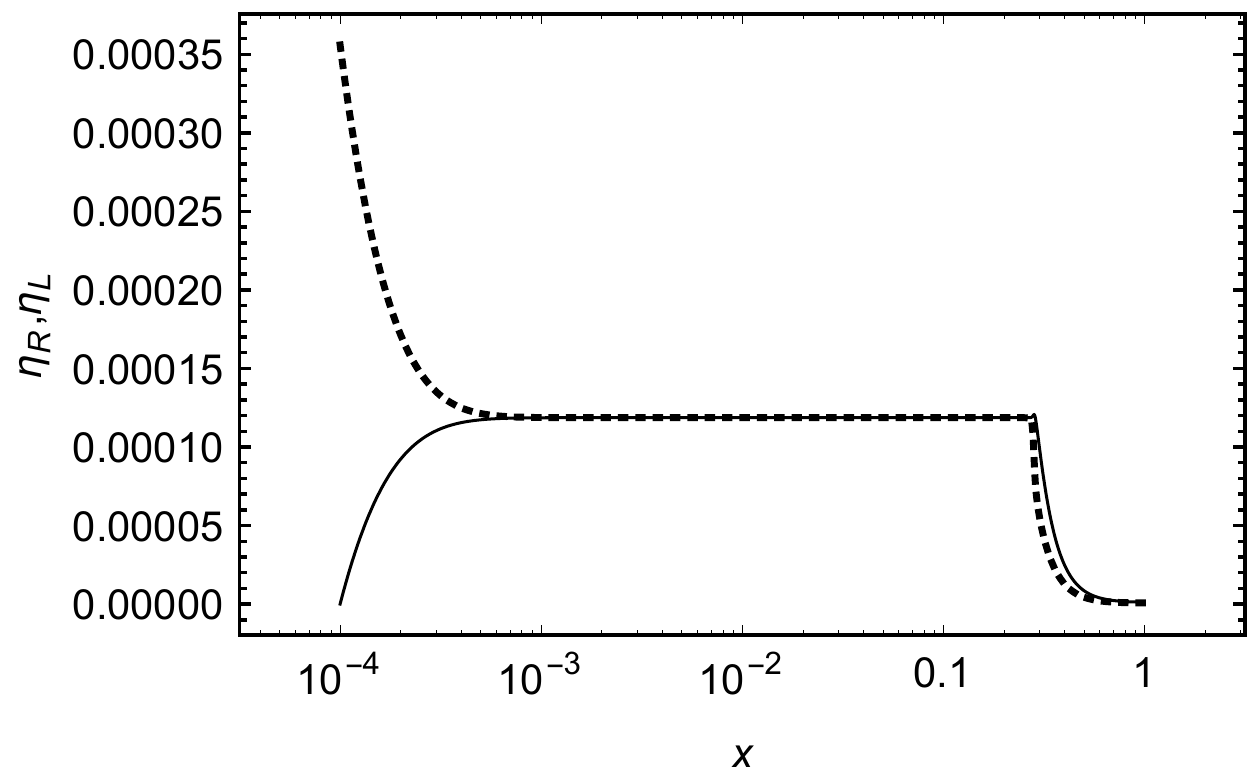}
  \hspace{2mm}
  \includegraphics[width=65mm]{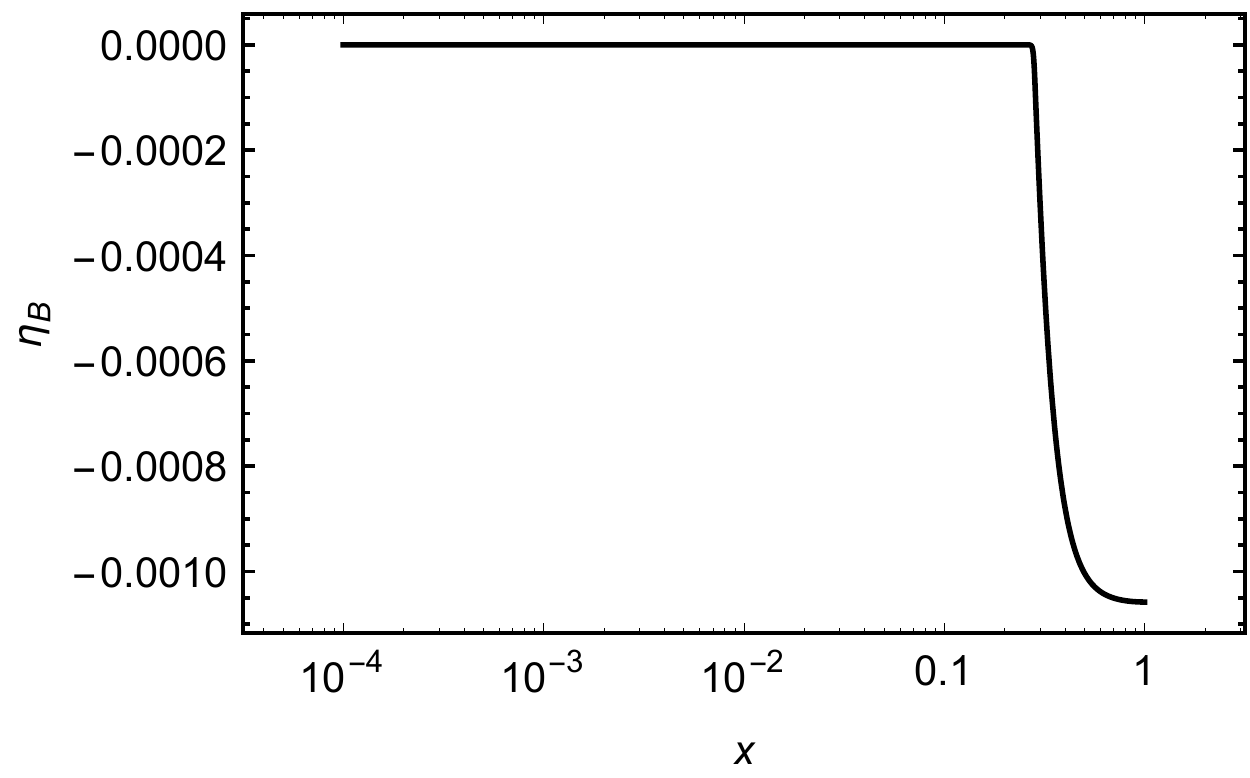}
\begin{center}
 \includegraphics[width=65mm]{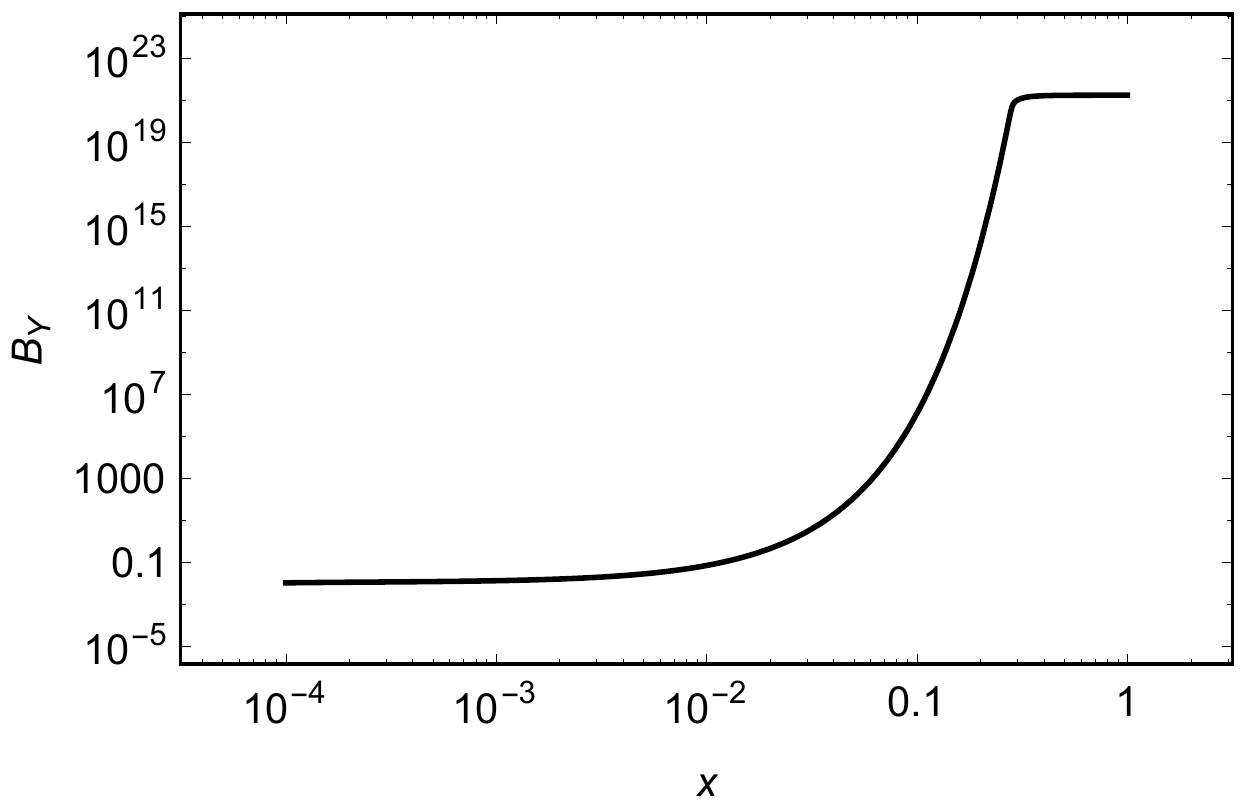}
\end{center}
\caption{The time plots of (top left): right-handed electron asymmetry $\eta_R=\eta_{e_R}$ (dotted line), left-handed lepton asymmetry $\eta_L=\eta_{e_L}=\eta_{\nu_e^L}$ (solid line), (top right): baryonic asymmetry and (bottom): hypermagnetic field amplitude, for $y_R^{(0)} = 10^3$, $k_0=10^{-7}T_{EW}$ and $B_Y^{(0)}=10^{-2}$G. The maximum relative error for these graphs is of the order of $10^{-13}$.} \label{three-timeplot}

\end{figure}

\begin{figure} 
  \includegraphics[width=65mm]{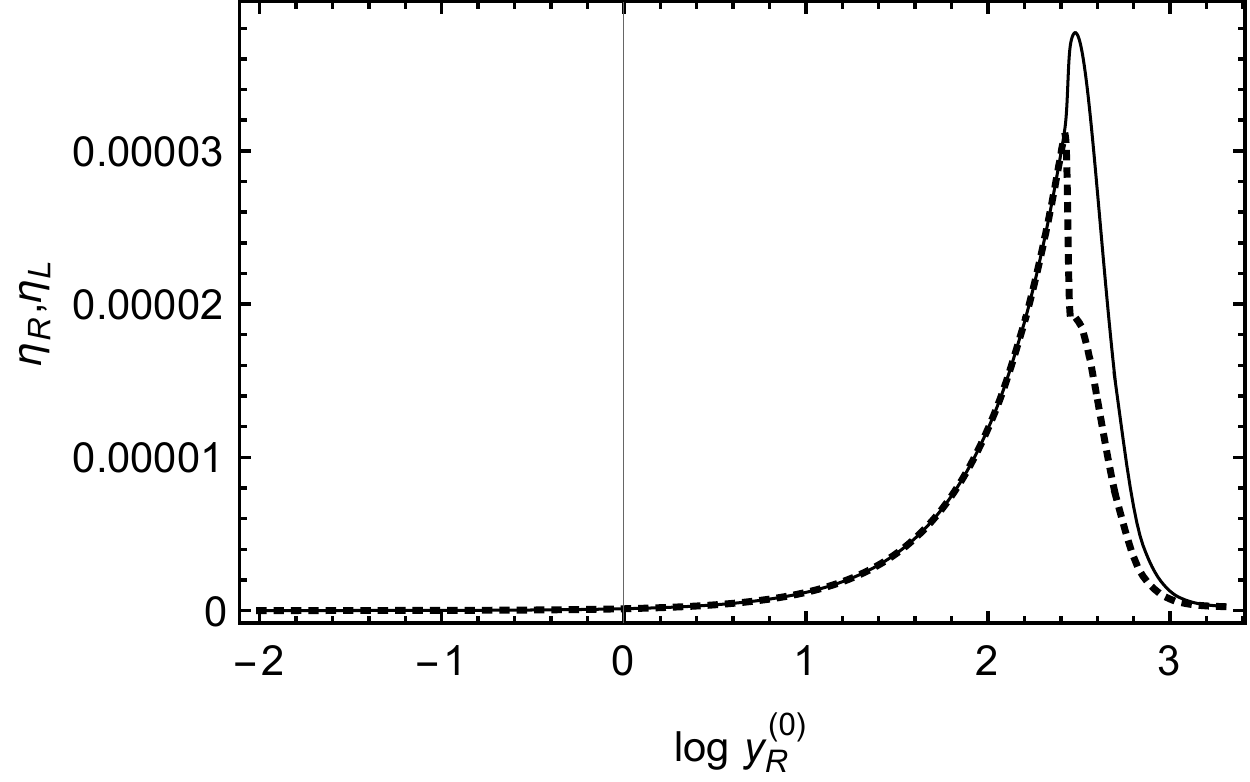}
  \hspace{2mm}
  \includegraphics[width=65mm]{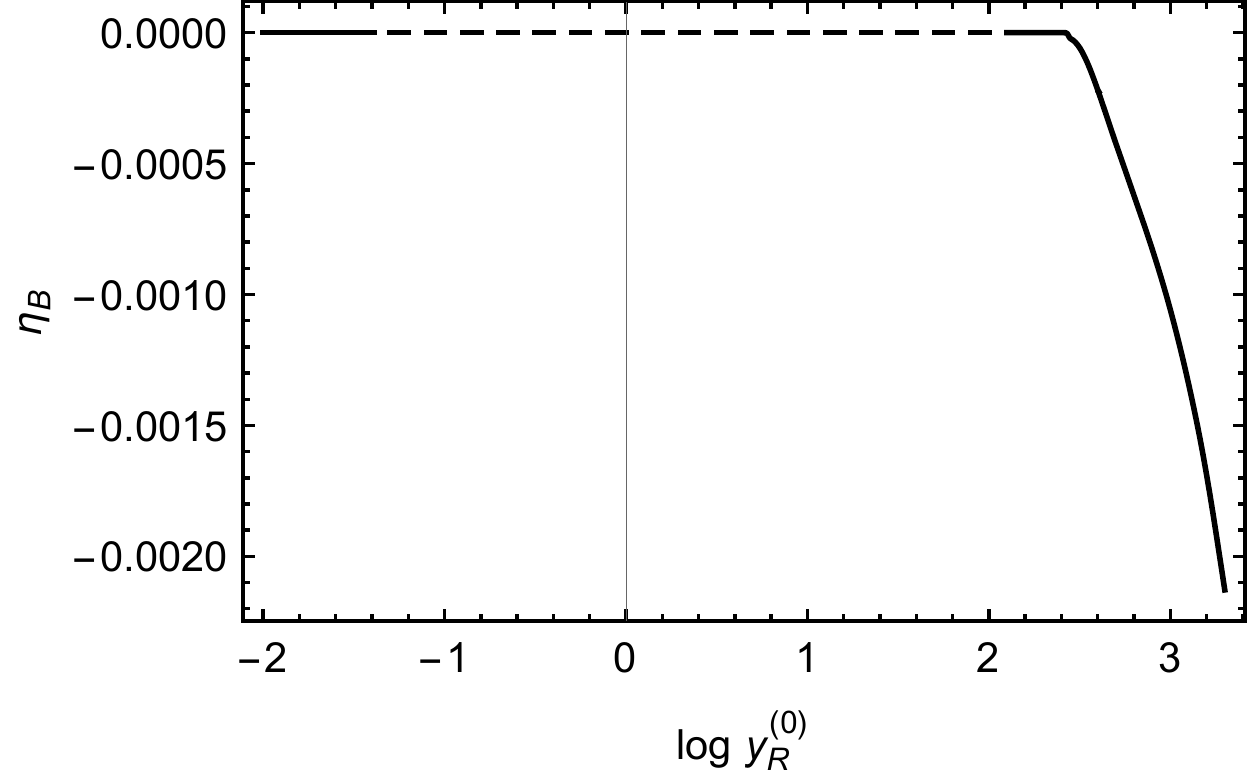}
\begin{center}
 \includegraphics[width=65mm]{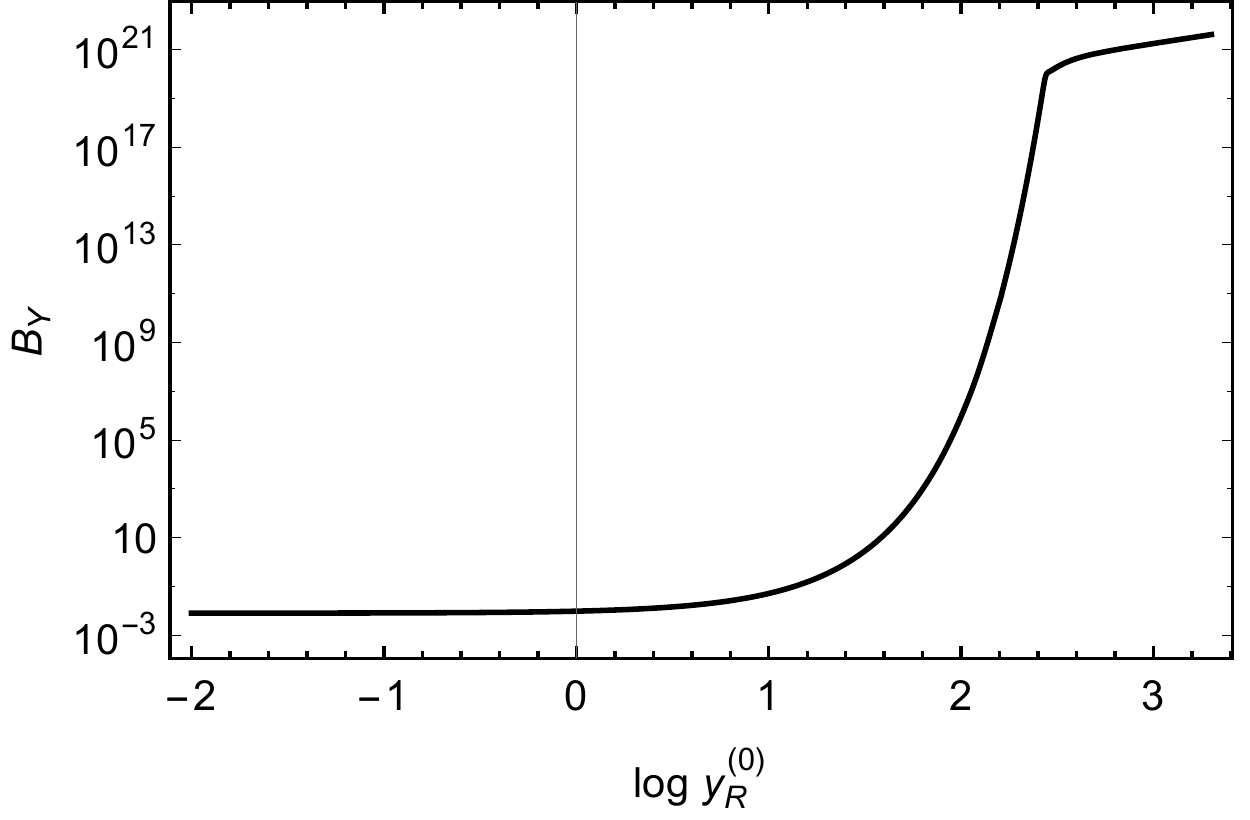}
\end{center}
\caption{(top left): The asymmetries of right-handed electrons $\eta_R=\eta_{e_R}$ (dotted line), left-handed leptons $\eta_L=\eta_{e_L}=\eta_{\nu_e^L}$ (solid line) at the EWPT time $t_{EW}$. (top right): The baryon asymmetry at the EWPT time $t_{EW}$. (bottom): The amplitude of hypermagnetic field at the EWPT time $t_{EW}$. It is assumed that $B_Y^{(0)}=10^{-2}$G, $k_0=10^{-7}T_{EW}$ and $y_R^{(0)}$ changes in the range $10^{-2}<y_R^{(0)}<10^{3.3}$. The maximum relative error for these graphs is of the order of $10^{-13}$.} \label{three-}

\end{figure}

The third part of our investigation has shown that when $B_Y^{(0)} = 0$, nothing interesting happens. Therefore, in the fourth part of our investigation we  examine the possibility of growing a seed of hypermagnetic field with a very small amplitude, e.g. $B_Y^{(0)}=10^{-2}$G, in the presence of a primordial right-handed electron asymmetry $y_R^{(0)} = 10^3$. The time plots are shown in Figure \ref{three-timeplot}. It is evident that the amplitude of hypermagnetic fields grows rapidly and gets as large as $10^{21}$G. In contrast, the leptonic asymmetries decrease nearly to zero and  the baryonic asymmetry drops to negative values. This is due to the fact that the initial value of $\eta_B/3-\eta_{L_e}$ is not zero. We then solve the evolution equations with the initial amplitude of hypermagnetic field $B_Y^{(0)}=10^{-2}$G and the normalized primordial right-handed electron asymmetry $y_R^{(0)}$ in the range $10^{-2}<y_R^{(0)}<10^{3.3}$. 
The final values of asymmetries and hypermagnetic field amplitude at the EWPT time $t_{EW}$ versus log of $y_R^{(0)}$ are presented in Figure \ref{three-}. It can be seen that, for $10^{1}<y_R^{(0)}<10^{2.4}$ the final value of the hypermagnetic field amplitude grows rapidly while, for $y_R^{(0)}>10^{2.4}$ the final value of asymmetries decrease with a greater slope and the hypermagnetic field amplitude at $t_{EW}$ grows with a much smaller slope. An interesting conclusion from these graphs is that $y_R^{(0)}=10^{2.4}$ is the most effective value for generating $B_Y(t_{EW})$ with minimal expenditure of matter asymmetries.

In the fifth and our final investigation, we solve the equations with our last assumptions but with the presence of initial baryonic asymmetry which fulfills the condition $\eta_B^{(0)}/3-\eta_{L_e}^{(0)}=0$. This prevents the baryonic asymmetry to drop to negative values. The results for the time plots are similar to those of the fourth part, except that $\eta_B$ approaches zero rather than a negative value. However, the behavior of $\eta_B$ at $t_{EW}$ changes completely and becomes similar to those of $\eta_R$ and $\eta_L$. The conclusion about $y_R^{(0)}=10^{2.4}$ still holds true here. 

\section{Summary and discussion}\label{Summary and discussion}
In this paper, we have studied the simultaneous evolution of matter asymmetries and hypermagnetic fields in the temperature range $100\mbox{GeV}<T<10\mbox{TeV}$. We have presented the effects of the sign change in the expression for $\alpha_Y$ within the context of the simple model introduced in Ref.\ \cite{Dvornikov2011}. We have found that matter asymmetry generation increases considerably and the hypermagnetic field is strengthened slightly as compared to the case studied in that reference (see Figure \ref{RGB}). For the rest of our analysis, we have used our proposed form for $\alpha_Y$ and set $k_0$ to $k_{max}=10^{-7}T_{EW}$. 

We have shown that matter asymmetry generation is possible via hypermagnetic fields. That is, a strong helical hypermagnetic field present in the plasma can produce and grow the asymmetries, however the growth process saturates for $B_Y^{(0)} \gtrsim 10^{21}$G (see Figure \ref{one-}). We have also shown that the amplification of hypermagnetic field is possible via matter asymmetries. Indeed, large matter asymmetries can grow a very weak seed of helical hypermagnetic field to an strong one (see Figure \ref{three-timeplot}). Nevertheless, they cannot generate any such hypermagnetic field when it has no seed in the plasma. 
Moreover, $y_R^{(0)}=10^{2.4}$ ($\eta_R^{(0)}\simeq8.94\times10^{-5}$) is an optimal initial value of right-handed electron asymmetry for obtaining large matter asymmetries and strong hypermagnetic field (see Figure \ref{three-}). 
Now, let us see what the observational data state about the baryon asymmetry of the Universe and the properties of the present magnetic fields, and whether our results are compatible with these observations.

The amount of baryon asymmetry of the Universe (BAU) is $\eta_B \sim 10^{-10}$, which has been determined independently in two different ways: from the abundances of light elements in the intergalactic medium (IGM), and from the power spectrum of temperature fluctuations in the cosmic microwave background (CMB) \cite{Canetti}. On the other hand, the observation of the CMB temperature anisotropy puts an upper bound on the current strength $B_0$ of magnetic fields as 
\be
B_0 \lesssim 10^{-9}\mbox{G},\ \ \ \ \ \ \ \ \ \ \ \ \ \ \ \ \ \mbox{for} \ \ \lambda_0 \gtrsim 1\mbox{Mpc},
\ee
where $\lambda_0$ is the current scale of the magnetic fields. Meanwhile the CMB distortion puts a slightly milder but nontrivial upper bound on $B_0$ on smaller scales. Moreover, the observations of the gamma rays from blazars put the following lower bounds on the strength $B_0$ of the present large scale magnetic fields \cite{Fujita}:
\be\begin{split}
B_0\gtrsim10^{-17}\mbox{G}{\left(\frac{\lambda_0}{1\mbox{Mpc}}\right)}^{-1/2}\ \ \ \ \ \ \mbox{for} \ \ \lambda_0<1\mbox{Mpc}\cr
B_0\gtrsim10^{-17}\mbox{G}\ \ \ \ \ \ \ \ \ \ \ \ \ \ \ \ \ \ \ \ \ \ \ \mbox{for} \ \ \lambda_0>1\mbox{Mpc} 
\end{split}\ee
Assuming that the time evolution of cosmic magnetic fields is trivial, i.e. an adiabatic evolution solely due to the cosmic expansion, their physical strength $B(t)$ and scale $\lambda(t)$ become proportional to $a^{-2}(t)$ and $a(t)$ respectively, where $a(t)$ is the scale factor. However, various effects can influence their time evolution, namely the interaction with turbulent fluid, the viscous diffusion, etc. Indeed, the MHD effects may cause the inverse cascade process, which is an important process for increasing the scale of magnetic fields, but needs large amounts of magnetic helicity to operate correctly. In this mechanism, the turbulence in the plasma may cause the magnetic correlation scale $\lambda(t)$ to grow faster than $a(t)$ \cite{Fujita}. Let us now evaluate the compatibility of our results with the observational data. 

Let us estimate roughly the scale of the hypermagnetic field used in our investigations via the relation $\lambda=k_0^{-1}$:
\be
\lambda_{(T=100\textrm{GeV})}=(10^{-7}T_{EW})^{-1}=10^5\mbox{Gev}^{-1}=2\times10^{-9}\mbox{cm}=6.45\times10^{-28}\mbox{pc},
\ee
which is a very small scale. Assuming that the time evolution of magnetic field from $T=100$GeV to $T=2\mbox{K}=17.2\times10^{-14}$GeV is trivial, we can compute the resultant scale of the magnetic field at present time via the mentioned relation $\lambda(t) \propto a(t) \propto T^{-1}$:
\be\begin{split}
\lambda_{(T=2\textrm{K})}=\lambda_{(T=100\textrm{GeV})}\left(\frac{\mbox{100GeV}}{17.2\times10^{-14}\mbox{GeV}}\right)\simeq3.75\times10^{-13}\mbox{pc}\cr
\end{split}\ee
Although the scale becomes larger, it is not within the acceptable range of present scales of magnetic fields. Let us decrease the wave number $k_0$ to
increase the scale $\lambda$ of our assumed configuration for the hypermagnetic field. This leads to a decrease of the matter asymmetries as well. When we use $k_0=10^{-4}k_{max}$, the maximum saturated value of the baryonic asymmetry in the left plot of Figure \ref{one-} becomes $\eta_B\simeq10^{-10}$. Therefore, this is the minimum $k_0$ which can give the baryonic asymmetry. This value of $k_0$ corresponds to $\lambda_{(T=100\textrm{GeV})}\simeq6.45\times10^{-24}\mbox{pc}$ which leads to $\lambda_{(T=2\textrm{K})}\simeq3.75\times10^{-9}\mbox{pc}$. We are still far from the current scales of magnetic fields. 

While the assumed scales of the hypermagnetic field ($10^7T_{EW}^{-1}$ or $10^{11}T_{EW}^{-1}$) are much larger than the mean distance between particles in plasma ($T^{-1}$), they are much less than the horizon size whose value at the EWPT is $l_H=M^*_{Pl}/T_{EW}^2=10^{16}/T_{EW}$, where $M^*_{Pl}$ is the reduced Planck mass. Since, the z axis direction of the hypermagnetic field is arbitrary, these fields are indeed small scale random fields that produce no anisotropy in the plasma. Therefore, it seems necessary to rely on an inverse cascade process starting after the EWPT for the generated Maxwellian magnetic fields to obtain the present large scale magnetic fields. 

 
Assuming that the inverse cascade process is the only nontrivial process operating after EWPT, and using the constraints coming from CMB and gamma rays from blazars, one obtains the relation $\lambda_0\simeq10^{-6}\mbox{Mpc}(\frac{B_0}{10^{-14}\textrm{G}})$ and the following ranges for $\lambda_0$ and $B_0$ \cite{Fujita}: 
\be
1\mbox{pc}<\lambda_0<1\mbox{Mpc},\ \ \ \ \ \ \ 10^{-14}\mbox{G}<B_0<10^{-8}\mbox{G}.
\ee

Let us assume the minimum $\lambda_0\simeq1$pc which corresponds to $B_0\simeq10^{-14}$G, and assume that the inverse cascade process starts immediately after the EWPT. Then, we can roughly estimate $\lambda$ and $B$ at $T=100$GeV by using the equations (2.4) and (2.5) from Ref.\ \cite{Fujita} to obtain
\be\begin{split}
B_{(T=100\textrm{GeV})}\simeq9.3\times10^{19}\mbox{G},\ \ \ \ \ \ \ \lambda_{(T=100\textrm{GeV})}\simeq 2.4\times 10^{-23}\mbox{pc}
\end{split}\ee
It is interesting to see that the above values for $B$ and $\lambda$ could be compatible with our results which are obtained using a simple model. That is, the approximate values of hypermagnetic field amplitude $B\sim10^{20}\mbox{G}$ and wave number $k_0\sim 10^{-4}k_{max}$ ($\lambda\sim 10^{-23}\mbox{pc}$) and $\eta_B\sim10^{-10}$ in our last investigation leads to the minimum values of amplitude and scale of present magnetic fields.\\

\noindent Acknowledgements: We would like to thank the research office of the Shahid Beheshti University for financial support.

\end{document}